\documentclass[twocolumn,amsmath,amssymb,prd,showpacs,superscriptaddress,preprintnumbers]{revtex4}

\usepackage{graphicx}
\usepackage{latexsym}
\usepackage{bm}

\newcommand{\beq}[1]{\begin{equation}\label{#1}}
\newcommand{\eeq}{\end{equation}}
\newcommand{\bea}[1]{\begin{eqnarray} \label{#1}}
\newcommand{\eea}{\end{eqnarray}}
\newcommand{\ba}{\begin{array}}
\newcommand{\ea}{\end{array}}

\newcommand{\rf}[1]{(\ref{#1})}

\newcommand{\lsim}{\mathrel{\vcenter{\hbox{$<$}\nointerlineskip\hbox{$\sim$}}}}
\newcommand{\gsim}{\mathrel{\vcenter{\hbox{$>$}\nointerlineskip\hbox{$\sim$}}}}

\newcommand{\unitmatrix}{\openone}
\newcommand{\half}{\frac{1}{2}}

\newcommand{\eps}{\epsilon}
\newcommand{\ra}{\rangle}
\newcommand{\la}{\langle}

\newcommand{\rarr}{\rightarrow}
\newcommand{\lrarr}{\leftrightarrow}
\newcommand{\order}{{\cal O}}

\newcommand{\Rearth}{R_\oplus}
\newcommand{\Rsun}{R_\odot}

\newcommand{\dmsq}{\delta m^2}

\def\U2{\underline{U\hspace{-.9mm}}\hspace{.9mm}}

\def\TEna{\tilde{E}_{N\!A}}
\def\Ena{E_{N\!A}}
\def\bbar{\bar b}
\def\nubar{\bar\nu}
\def\nue{{\nu_{e}}}
\def\numu{{\nu_{\mu}}}
\def\nutau{{\nu_{\tau}}}
\def\nuebar{{\bar \nu}_e}
\def\numubar{{\bar \nu}_\mu}
\def\nutaubar{{\bar \nu}_\tau}

\newcommand{\we}{w_e}
\newcommand{\wmu}{w_\mu}
\newcommand{\wtau}{w_\tau}
\newcommand{\walpha}{w_\alpha}
\newcommand{\webar}{w_{\bar e}}
\newcommand{\wmubar}{w_{\bar \mu}}
\newcommand{\wtaubar}{w_{\bar \tau}}
\newcommand{\walphabar}{w_{\bar \alpha}}
\newcommand{\rhoearth}{\rho({\rm Earth})}

\begin{document}

\preprint{MIT-CTP-3849}
\preprint{MPP-2007-97}

\title{Neutrino flavor ratios as diagnostic of solar WIMP annihilation}

\author{Ralf Lehnert}

\email[Electronic mail: ]{rlehnert@lns.mit.edu}
\affiliation{Center for Theoretical Physics,
Massachusetts Institute of Technology, Cambridge, MA 02139, USA}
\affiliation{Max--Planck--Institut f\"ur Physik, F\"ohringer Ring 6, 80805 M\"unchen, Germany}

\author{Thomas J. Weiler}
\email[Electronic mail: ]{tom.weiler@vanderbilt.edu}
\affiliation{Department of Physics and Astronomy,
Vanderbilt University, Nashville, TN 37235, USA}

\date{\today}

\begin{abstract} 
We consider the neutrino (and antineutrino) flavors arriving at Earth 
for neutrinos produced in the 
annihilation of weakly interacting massive particles (WIMPs) in the Sun's core. 
Solar-matter effects on the flavor propagation of the resulting $\agt$~GeV neutrinos 
are studied analytically within a density-matrix formalism. 
Matter effects, including mass-state level-crossings, 
influence the flavor fluxes considerably. 
The exposition herein is somewhat pedagogical, 
in that it starts with adiabatic evolution of single flavors from the Sun's center,
with $\theta_{13}$ set to zero,
and progresses to fully realistic processing of the flavor ratios expected in WIMP decay,
from the Sun's core to the Earth. 
In the fully realistic calculation, non-adiabatic level-crossing is included, 
as are possible nonzero values for $\theta_{13}$ 
and the CP-violating phase $\delta$.
Due to resonance enhancement in matter,
nonzero values of $\theta_{13}$ even smaller than a degree 
can noticeably affect flavor propagation.
Both normal and inverted neutrino-mass hierarchies are considered.
Our main conclusion is that measuring flavor ratios (in addition to energy spectra) 
of $\agt$~GeV solar neutrinos
can provide discrinination between WIMP models.
In particular, we demonstrate the flavor differences at Earth for neutrinos from 
the two main classes of WIMP final states, namely $W^+ W^-$ and 
95\% $b\,\overline{b}$ + 5\% $\tau^+\tau^-$.
Conversely, if WIMP properties were to be learned from production in 
future accelerators, then the flavor ratios of $\agt$~GeV solar neutrinos might be useful for 
inferring $\theta_{13}$ and the mass hierarchy.
From the full calculations,
we find (and prove) some general features:
a flavor-democratic flux produced at the Sun's core arrives at earth still flavor-democratic;
for maximal $\theta_{32}$ but arbitrary $\theta_{21}$ and $\theta_{13}$, 
the replacement $\delta\rarr\pi-\delta$ leaves the $e$~flavor spectra unaltered 
but interchanges $\mu$ and $\tau$~spectra at Earth; and,
only for neutrinos in the inverted hierarchy 
and antineutrinos in the normal hierarchy
is the dependence on the mixing phase $\delta$ not suppressed to order
$\delta m^2_{21}/\delta m^2_{32}$. 
\end{abstract}


\maketitle

\section{Introduction}
\label{sec:intro}
One of the most fundamental questions about our Universe
is its matter content.  Recent cosmological observations
of type Ia supernovae~\cite{sn},
the cosmic microwave background~\cite{cmb},
galaxy-cluster evolution~\cite{galaxy}, 
and gravitational lensing~\cite{lens}
show 
that known particle species contribute only 
$\Omega_{\rm B}\simeq$~4\% 
to the total energy density of the Cosmos.
Abundance measurements of light elements 
together with the theory of Big-Bang Nucleosynthesis
largely support these findings~\cite{bbn}.
The above results also indicate
the presence of two---as yet unknown---matter components:
``dark energy" and ``dark matter"
with relative densities of 
$\Omega_{\rm DE}\simeq$~73\% and $\Omega_{\rm DM}\simeq$~23\%, 
respectively.
Unraveling the nature of these novel types of matter 
lies currently on the forefront of fundamental-physics research.

One of the new components, 
dark matter (DM), 
exhibits gravitational clustering 
and is therefore believed to be composed of novel massive particles.
Further compelling support for this idea 
is provided by theoretical approaches
to physics beyond the Standard Model,
which typically require additional particles for consistency. 
Weakly interacting massive particles (WIMPs)
are a general and popular candidate in this context. 
A particularly attractive example of a WIMP 
is the lightest supersymmetric particle (LSP),
usually the neutralino, 
present in various SUSY extensions of the Standard Model~\cite{JKGriest}. 
This particle is a mixture of the four superpartners of the two neutral Higgs particles 
and the two neutral electroweak gauge bosons.  
The neutralino is expected to have a mass of order $10^2$ GeV.
The attractiveness of identifying the LSP with dark matter is twofold:
First, the LSP is stable, assuming $R$-parity is an unbroken symmetry.
$R$-parity is introduced to stabilize the proton to the level required by experiment.
Second, a standard thermodynamic calculation of decoupling yields the LSP abundance 
to be $\order (10^{-34}{\rm cm}^2/\sigma_A)$, where $\sigma_A$ is the WIMP 
annihilation cross section for Majorana particles like the neutralino 
(for Dirac WIMPS, there is an additional factor of 
$T_{\rm decouple} /M_{\rm WIMP}\sim 1/30$).
Inputting the usual weak cross section 
$\sigma_{\rm A}(s=M_{\rm WIMP}^2) \sim G_F^2/\pi$,
one naturally explains the inferred value $\Omega_{\rm DM}\simeq 23\%$~\cite{lund,Note3}. 

Besides direct detection via WIMP scattering in cryogenic detectors,
indirect searches for WIMP annihilation into gamma rays, antimatter, and neutrinos
are currently being pursued 
as promising experimental avenues~\cite{BHSilk,HooperTaylor}.
The gravitational fields of the Sun and the Earth 
can capture large numbers of WIMPs,
which are expected to then infall toward the center of the body~\cite{dm_trap}.
Since the annihilation rate is proportional to $n_{\rm DM}^2$,
annihilations of WIMPs captured by the Sun or Earth may be considerable. 
An important factor in WIMP capture is the surface area of the capturing body.
Another is the efficiency of capture $\eps_{\rm cptr}$, 
which is related to the depth of the potential well
relative to the mean kinetic energy of the infalling WIMP.
The capture rate is proportional to the square of the body area
and to the capture efficiency.
There results a relative Sun-to-Earth capture rate of 
$(\eps^{\rm Sun}_{\rm cptr}/\eps^{\rm Earth}_{\rm cptr})\,
(\Rsun/\Rearth)^2$,
where $\Rsun=6.96 \times 10^{10}$~cm is the radius of the Sun
and $\Rearth=6.38\times 10^8$~cm the radius of the Earth.
Allowing for the $1/D^2$ flux dilution between the source and the earthly 
detector, one then arrives at a na\"ive relative 
flux ratio at Earth of 
$F_{\rm Sun}/F_{\rm Earth}\sim
 (\eps^{\rm Sun}_{\rm cptr}/\eps^{\rm Earth}_{\rm cptr})\,
 (\Rsun/\Rearth)^2\,(\Rearth/{\rm AU})^2 = 
 2\times 10^{-5}\,(\eps^{\rm Sun}_{\rm cptr}/\eps^{\rm Earth}_{\rm cptr})$.
Three further pieces of physics more than compensate the $10^{-5}$ factor 
and make the Sun the better source for experimental study.
First, 
the capture efficiency of the Sun far exceeds that of the Earth,
with relative potentials~$\sim M/R$~that are 3000 times larger for the Sun.
The second is that the solar potential reduces the lifetime of WIMP orbits among the planets,
thereby reducing the capture rate at Earth~\cite{lund}.
The third factor is 
that center-pointing gravitational force $\sim M(<r)/r^2$ is far greater
for the Sun, due to its roughly exponential relation between density and radius.
In fact, the DM in the Earth is not concentrated in Earth's center
but rather broadly distributed in $r$.
As a consequence, $n_{\rm DM}^2$ is not optimized in the Earth, and the annihilation signal is fairly 
diffuse because it comes from a large fraction of a steradian.
On the other hand, the Sun has a high central DM density, and an annihilation signal coming from 
a fraction of a square degree of solid angle.
Thus, we will not consider further the WIMP annihilation flux coming from the Earth,
but rather focus on the annihilation signal from the Sun.

The solar medium absorbs all annihilation products except neutrinos
with energies below tens of GeV.
The typical energies of fusion neutrinos are 10~MeV or less,
while the typical energies of neutrinos and antineutrinos 
from the WIMP decay chain are GeV to tens of GeV.
It follows then that the detection of higher-energy neutrinos from the Sun
provides an excellent tool for indirect WIMP detection.

The idea of inferring WIMPs by measuring the associated high-energy 
neutrino flux from the Sun's center
has been the focus of numerous theoretical studies~\cite{noMSW},
and first experimental results have placed loose constraints
on such scenarios~\cite{kamiokande}.
Note, however, that neutrino detection is highly dependent on neutrino flavor
($\nu_e$ versus $\numu$ versus $\nutau$).
These flavors evolve as the neutrinos travel in space and time.
This is particularly so in the Sun, where 
the solar environment adds significant  matter-dependent effects 
to the flavor evolution~\cite{MSW}.
In fact, matter effects do not only alter the flavor evolution, 
but they also treat the neutrino and antineutrino differently. 
Therefore, dark-matter searches via annihilation neutrinos typically require considerable 
knowledge of neutrino mixing and mass parameters.  

Matter effects were not included in the early works cited in~\cite{noMSW}.
More recently, the MSW effect has been incorporated 
into the analysis of solar WIMP annihilation neutrinos 
by a number of authors~\cite{ellis92,gouvea92,crotty02,Cirelli05}.
Of these, only~\cite{Cirelli05} is recent enough to have focused on the 
``last-solar-model-standing,'' the large mixing angle (LMA) solution.
The other papers hedged their bets among LMA, SMA, LOW, and VAC solutions
appropriate for their time when less was known~\cite{advertisement}. 

Although our knowledge of neutrino parameters has advanced quickly,
it remains incomplete. 
The neutrino mixing angles seem to converge to a ``tribimaximal'' form~\cite{tribi} 
with values in vacuum given to a good approximation by 
$\sin^2\theta_{23}\sim 1/2$, $\sin^2\theta_{12}\sim 1/3$, 
and $\sin^2\theta_{13}\sim 0$. 
The large value of the ``solar'' angle $\theta_{12}$ lies at the heart of the LMA solution. 
The neutrino mass-squared differences have converged to 
$\dmsq_{23}\sim 2.5\times 10^{-3}{\rm eV}^2$ and 
$\dmsq_{12}\sim 0.8\times 10^{-4}{\rm eV}^2$. 

In Ref.~\cite{Cirelli05}, numerically precise terrestrial spectra
resulting from the primary WIMP-annihilation channels are predicted, 
signal topologies in neutrino telescopes are given,
and the possibility of discriminating among dark-matter models is discussed.
Our investigation complements that in Ref.~\cite{Cirelli05} mainly in two ways.

First,
we focus on particularly promising observables:
there has been a growing awareness in recent years that
neutrino telescopes can infer {\em flavor ratios} of arriving neutrinos.
It has been shown~\cite{BBHPWratios}
that neutrino flavor ratios
may be determined in ice and water neutrino telescopes
from the observation of the ratio of muon-track events
to jet events,
where the former arise from charged-current (CC) $\numu$ scattering,
and the latter from $\nue$ and $\nutau$ CC events and all neutral-current (NC) events. 
The muon mean-free-path (MFP) 
before decay is $c\tau_\mu\sim 6.3\,{\rm km}\,(E_\mu/{\rm GeV})$,
which allows even 100~MeV muons to be identified. 
Large-volume magnetized-iron calorimeters, liquid-Argon detectors,
and active-scintillator detectors have the potential
to perform even better neutrino-flavor identification~\cite{BCHHWliqar}.
The $\tau$ MFP is $c\tau_\tau\sim 0.49\times 10^{-2}\,{\rm cm}\,(E_\tau/{\rm GeV})$,
leading to possibly identifiable $\tau$ tracks for energies above 10~GeV.

Second,
we use a related but different treatment of the propagation of $E\gsim$~
GeV neutrinos from the solar core to Earth.
We employ a density-matrix formalism and 
arrive at compact and convenient analytic results
which we believe improve our intuition. 

Besides presenting matter effects that alter the flavor content of a 
propagating neutrino flux, possibly resonantly,
solar matter also absorbs neutrinos above some energy $E_{\rm abs}$,
due to the neutrino inelastic CC and NC interactions.
We wish to neglect absorption 
by looking at neutrinos with $E < E_{\rm abs}$.
So we must estimate $E_{\rm abs}$.
In the Sun,
the number density of electrons $n_e$ decreases nearly exponentially 
with the distance $r$ from the Sun's center.
An approximate expression for the solar density profile $n_e(r)$
is given by~\cite{solar_model}
\beq{number_density}
n_e(r)=245\;N_A\;e^{-r/\lambda_{\odot}}\;{\rm cm}^{-3}\,,
      \quad \lambda_\odot = 0.095\,R_\odot\,,
\eeq
for $r\leq R_{\odot}$; 
$N_A$ is Avogadro's number.
Equation \rf{number_density} also serves 
as a rough estimate of the proton and neutron densities.

Integration of this expression from $r=0$ to $r=R_\odot$
gives the column density $S_j$ 
from the Sun's center to its surface:
\beq{column_density}
S_e=S_p\simeq S_n \simeq 1.6\,N_A\,\times 10^{12}\,{\rm cm}^{-2}\,.
\eeq
(For $R_\odot \gg \lambda_\odot$, one also has $S_j\simeq n_j(0)\,\lambda_\odot$.)
The neutrino's optical depth $\tau_\nu$ is the 
number of scattering MFPs from the Sun's center to the surface
(the most probable number of interactions in transit);
the fraction of produced neutrinos that escape the Sun is $e^{-\tau_\nu}$.
The optical depth is equal to the product of cross section $\sigma_\nu$ 
and column density.
Since $\sigma_{\nu N}\gg \sigma_{\nu e}$
and $S_p+S_n\simeq 2\,S_e$, we have for the optical depth
$\tau_\nu\sim 2\,\sigma_{\nu N}\,S_e$.
If we ask that a fraction $f$ or more of the neutrino flux escapes from the Sun,
we require 
\beq{optical_depth}
\tau_\nu \le \ln f^{-1}\,.
\eeq
With the onset of deep inelastic scattering at $\sim$~GeV energies,
the neutrino and antineutrino cross sections grow linearly with $E$
until $W$-propagator effects become important near $E\sim M_W^2/2\,m_N\sim$~TeV.
To a good approximation, the expression 
for the inelastic CC+NC neutrino and antineutrino cross sections 
at energies between 1 GeV and 1 TeV is~\cite{cross_sec}:
\beq{cross_sec_estimate}
\sigma_{\nu N}(E)\simeq 2\,\sigma_{\nubar N}(E) 
\simeq\frac{E}{\rm GeV}10^{-38}\,{\rm cm}^2\, .
\eeq
The rules of thumb are that the neutrino cross section is twice the antineutrino 
cross-section, and the CC contribution is a bit over twice the NC contribution.
Also, the fractional energy loss per scatter is about 1/2 for neutrinos
at the energies relevant here, 
and half that again for antineutrinos.  
In CC scattering, the neutrinos and antineutrinos 
are lost upon the first scatter; in NC scattering, the neutrinos and antineutrinos
persevere but with the fractional energy losses just presented.
Since the NC/CC scattering ratio is less than a half, 
we may and will neglect NC scattering in what follows.

The energy $E_{\rm abs}$ can now be calculated by
substituting Eq.~\rf{cross_sec_estimate} and twice times Eq.~\rf{column_density}
for $\tau_\nu$ in Eq.~\rf{optical_depth}.
The result is $E_{\rm abs}=50\,(\ln f^{-1})\,{\rm GeV}$ for neutrinos,
and twice that for antineutrinos.
Some explicit numerical results for $E^{\nubar}_{\rm abs}=2\,E^{\nu}_{\rm abs}$ are 
$10$~GeV when $f=90\%$,
$36$~GeV when $f=70\%$, 
and $70$~GeV when $f=50\%$.
If we discount the NC processing of neutrinos and attribute absorption to 
only the CC, then $E_{\rm abs}$ is 50\% larger.
A careful numerical calculation of absorption versus energy is given 
in Fig.\ 1 of Cirelli {\em et al.}~\cite{Cirelli05}.
The conclusion there and here is that 
up to energies of $\sim 10$~GeV,
absorption in the Sun can be neglected,
and above 100 GeV absorption is severe.

The format of this paper is as follows.
Flavor oscillations are presented in Sec.~\ref{sec:basics}.  
Matter effects, both adiabatic and non-adiabatic, 
the two possible mass hierarchies,
and zero versus nonzero $\theta_{13}$ and $\delta$ 
are discussed in detail.
In particular, two interesting theorems for the $\delta$ dependence of the neutrino flavors, 
valid for the adiabatic regime, 
are presented.
The application of the formalism to the solar WIMP annihilation is made in Sec.~\ref{sec:models}.
A summary and conclusions comprise Sec.~\ref{sec:sumNconclude}.
Some technical details are collected in four appendices.

\section{Physics of Flavor Mixing}
\label{sec:basics}

This section presents a density-matrix approach
for the flavor evolution
of high-energy neutrinos
injected at the center of the Sun.
In the first two subsections, 
certain simplifying assumptions
regarding neutrino-mixing parameters
and neutrino propagation in matter are employed;
our goal is 
to set up our formalism and notation, 
to provide intuition about the involved effects,
and to set the stage for a more sophisticated study
in the third subsection.

\subsection{Review: oscillations without matter}
\label{subsec:vacuumoscillations}
In the absence of a matter background
(i.e., in vacuum),
the free-neutrino Hamiltonian is diagonal in the mass basis:
\beq{Hmassvacuum}
H_M = E\,\unitmatrix +\frac{1}{2\,E}\,{\rm diag}(m^2_1, m^2_2, m^2_3)+{\cal O}(m^4/E^3)\,.
\eeq
Here, $E$ is the neutrino's energy,
and $\unitmatrix$ is the $3\times3$ unit matrix.
We label the mass eigenvectors in vacuum as $|j\ra = |1\ra$, $|2\ra$, $|3\ra$.
However, 
neutrinos are produced by the weak interaction.
In this production basis, 
called the ``weak basis'' or ``flavor basis,''
the interaction eigenstates are labeled by the flavors of their charged-current partners, 
as $|\alpha\ra = |e\ra$, $|\mu\ra$, $|\tau\ra$.
In general, 
the flavor basis is rotated with respect to the mass basis,
which leads to periodic flavor oscillations among propagating neutrinos.
The oscillation length is given by 
\bea{vaclength}
\lambda_{\rm V} &=& \frac{4\,\pi\,E}{\dmsq_{jk}}\nonumber\\
                &=& 0.036\,\left(\frac{E}{\rm GeV}\right)\,
                    \left(\frac{10^{-4}{\rm eV}^2}{\dmsq_{jk}}\right)\,R_\odot\,,
\eea
where $\dmsq_{jk}\equiv m_j^2-m_k^2$
denotes the neutrino mass-squared differences.
As mentioned in Sec.~\ref{sec:intro},
the values of $\dmsq_{jk}$ inferred from solar and atmospheric neutrino-oscillation experiments are
$\dmsq_{21}\sim 10^{-4}\,{\rm eV}^2$ and $\dmsq_{32}\sim 10^{-3}\,{\rm eV}^2$,
respectively.
Thus, the vacuum oscillation lengths are well contained within the Sun
for energies below 30~GeV.

The transformation between the mass basis 
and the flavor basis
is described by the vacuum mixing matrix
$U_{\alpha j}=\la j |\alpha\ra =\la \alpha |j\ra ^*$ 
with $\alpha = e,\,\mu,\,\tau$ and $j=1,\,2,\,3$.
The conventional parametrization of the vacuum mixing matrix invokes 
three angles and a phase~\cite{pdg}:
\beq{vacPDG}
U= R_{23}(\theta_{23})\,U^{\dagger}_{\delta}\,
  R_{13}(\theta_{13})\,U_{\delta}\,R_{12}(\theta_{12})\,,
\eeq
where $R_{jk}(\theta_{jk})$
determines a rotation in the $jk$ plane
by an angle $\theta_{jk}$.
The phase matrix
\beq{phasematrix}
U_{\delta}  = 
\left(\begin{array}{ccc}
e^{i\delta/2} & 0 & 0\\ 
0 & 1 & 0\\
0 & 0 & e^{-i\delta/2}
\end{array}\right)
\eeq
may be thought of as complexifying $R_{13}(\theta_{13})$ to
\bea{R13complex}
R_{13}(\theta_{13},\delta)&\equiv& 
U^{\dagger}_{\delta}\,R_{13}\,U_{\delta}\nonumber\\
  &=& \left(\begin{array}{ccc}
\cos\theta_{13} & 0 & e^{-i\delta}\sin\theta_{13}\\
0 & 1 & 0\\
-e^{+i\delta}\sin\theta_{13} & 0 & \cos\theta_{13}
\end{array}\right)\,.
\eea
We have omitted two additional Majorana phases,
as they leave unaffected neutrino oscillations.

As mentioned before, 
the central values of the mixing angles 
inferred from oscillation experiments 
are quite consistent with the tribimaximal angles~\cite{tribi} 
given 
explicitly by 
$\theta_{32}=45^\circ$, 
$\theta_{21}=35.26\dots^\circ$, 
and $\theta_{13}=0$.
The full tribimaximal mixing matrix is
\beq{tribimax}
U = 
\left(\begin{array}{rrc}
 \sqrt{\frac{2}{3}}  &  \sqrt{\frac{1}{3}} &  0                  \\
-\sqrt{\frac{1}{6}}  &  \sqrt{\frac{1}{3}} &  \sqrt{\frac{1}{2}} \\
-\sqrt{\frac{1}{6}}  &  \sqrt{\frac{1}{3}} & -\sqrt{\frac{1}{2}}
\end{array}\right)\,.
\eeq
We remind the reader 
that rows are labeled from top to bottom 
by the flavor indices $e$, $\mu$, and $\tau$,
and columns are labeled left to right by mass-eigenstate indices 1, 2, and~3.

Due to decoherence of the propagating neutrino phases, 
we shall see 
that it is the classical probabilities  $|U_{\alpha j}|^2$,
obtained by squaring the mixing-matrix elements,
that are of interest to us.
We collect these squared elements into a matrix 
that we denote as $\U2$.
For tribimaximal mixing,
its explicit form is 
\beq{tribimaxsq}
\U2 \equiv \frac{1}{6}\,
\left(\begin{array}{ccc}
 4 & 2 & 0   \\
 1 & 2 & 3 \\
 1 & 2 & 3
\end{array}\right)\,.
\eeq
One cannot avoid noticing the simplicity of the tribimaximal probabilities:
the $\nu_e$ contents of $\nu_1$, $\nu_2$, and $\nu_3$ are
2/3, 1/3, and 0, respectively, 
with the remaining probability being shared equally between $\nu_\mu$ and $\nu_\tau$.

The 3-$\sigma$ experimental limits on the mixing-angle ranges are~\cite{3sigma}
\beq{theta32limit}
0.34\leq\sin^2\theta_{32}\leq 0.68\,,
\eeq
\beq{theta21limit}
0.24\leq\sin^2\theta_{21}\leq 0.41\,,
\eeq
\beq{theta13limit}
\sin^2\theta_{13}\leq 0.044,\ \ {\rm or}\ \ \theta_{13}\leq 12^\circ\,.
\eeq
Notice that with $\U2_{e3}=\sin^2\theta_{13}$,
the ``zero'' in $\U2$ may be as large as 4\%.
Later in this paper we will explore the sensitivity of our results to 
experimentally allowed variations in $\theta_{13}$.

Neglecting matter effects, 
the density matrix for neutrino production in the Sun is 
\bea{rhovacuum1}
\rho &=& \we\,|e\ra\la e|+\wmu\,|\mu\ra\la\mu|+\wtau\,|\tau\ra\la\tau|\nonumber\\
     &=& \sum_\alpha w_\alpha\,U^*_{\alpha j}\,U_{\alpha k}\,|j\ra\la k|\,,
\eea
where the weights $w_\alpha$ are the relative flavor ratios at production 
normalized to obey $w_e+w_\mu+w_\tau=1$.
The density matrix for antineutrino production is given by the 
same formula when the $w_\alpha$ are replaced with antineutrino 
production weights $w_{\bar \alpha}$.
The first and second expressions in Eq.~\rf{rhovacuum1} 
present the production density matrix in the flavor and mass bases, respectively.
The point of exhibiting $\rho$ in the mass basis is to twofold.
First,
this basis diagonalizes the free-neutrino Hamiltonian
and is therefore appropriate for describing neutrino propagation to Earth.
Second,
it facilitates a discussion of decoherence,
as is explained next.

As the neutrinos propagate from the Sun to the Earth,
each off-diagonal operator $|j\ra\la k|$
acquires a phase factor $\exp(-i\phi_{kj})$ 
with large phase $\phi_{kj}=L\,\dmsq_{kj}/2E$,
where $L$ is the distance from neutrino-production site to the 
detection site at Earth.
The size of the Sun's DM core
and the detector's size 
contribute a large $\delta L$, 
and energy resolution in the detector contributes a $\delta E$
to the variance in the phase.
For a statistical sample of events, 
these uncertainties randomize the phases, 
effectively equating the phase factor 
to its mean value $\la\;\exp(-i\phi_{kl})\;\ra =0$.
Thus, 
off-diagonal elements of $\rho$, 
associated with the initial coherence of the neutrino wave function, 
may be dropped.
The loss of phase information allows us 
to work with classical probabilities.
We effectively obtain 
\beq{rhovacuum2}
\rhoearth = \sum_\alpha\,w_\alpha\,|U_{\alpha j}|^2\,|j\ra\la j|
\eeq
for the density matrix 
describing neutrinos arriving at Earth.

With the above result,
the flavor probabilities at Earth are given by
\beq{probvacuum}
P_{\nu_\odot\rarr\nu_\beta} = \la\beta|\rhoearth|\beta\ra =
\sum_{\alpha,j} w_\alpha\,|U_{\alpha j}|^2\,|U_{\beta j}|^2 \,.
\eeq
In matrix form,
this equation becomes 
\beq{matrixform}
\left(
\ba{l}
P_{\nu_\odot\rarr\nu_e}\\
P_{\nu_\odot\rarr\numu}\\
P_{\nu_\odot\rarr\nutau}
\ea
\right)
 = \U2\,\U2^T
\left(
\ba{l}
\we\\
\wmu\\
\wtau
\ea
\right)\,.
\eeq
Inputting the tribimaximal values leads to the explicit ``flavor-propagation matrix''
\beq{FPtribimax}
\U2\,\U2^{T} = \frac{1}{18}
\left(
\begin{array}{ccc}
 10 & 4 & 4 \\
 4 & 7 & 7  \\
 4 & 7 & 7  
\end{array}
\right)\,,
\eeq
and to 
\bea{}
\label{probvacuum2a}
P_{\nu_\odot\rarr\nu_e} &=& \frac{1}{18}\, \left[ 10\,w_e+4\,(\wmu+\wtau ) \right] \nonumber\\
         &=& \frac{1}{9}\;(2+3\,w_e)\,,\\
\label{probvacuum2b}
P_{\nu_\odot\rarr\nu_\mu} &=& P_{\nu_\odot\rarr\nu_\tau}
         = \frac{1}{18}\,\left[4\,\we+7\,(\wmu+\wtau )\right]\nonumber\\ 
        &=& \frac{1}{18}\;(7-3\,w_e)\,.
\eea
The equations for antineutrino flavor probabilities are obtained 
from these neutrino results via the replacement 
$w_\alpha \rightarrow w_{\bar \alpha}$.

In Eq.~(\ref{probvacuum2b}), 
the two probabilities 
$P_{\nu_\odot\rarr\nu_\mu}$ and $P_{\nu_\odot\rarr\nu_\tau}$
are equal.  
Also, in the first expressions on the right 
in Eqs.~(\ref{probvacuum2a}) and (\ref{probvacuum2b}),
$\wmu$ and $\wtau$ enter symmetrically.
These results are manifestations of a $\numu$--$\nutau$ 
interchange symmetry inherent in the tribimaximal mixing matrix.
The $\numu$--$\nutau$ interchange symmetry is exact for 
$\theta_{23}= 45^\circ$ (maximal $\numu$--$\nutau$ mixing) 
and $\theta_{13}=0$~\cite{Note4}.
The interchange symmetry holds in matter as well as in vacuum 
because the $\numu$--$\nutau$ sector remains unaffected by the matter potential.
More manifestations of the interchange symmetry will become evident in 
expressions and figures to be presented in subsequent sections.
In the second expressions on the right in Eqs.~\rf{probvacuum2a} and \rf{probvacuum2b},
$\we+\wmu+\wtau=1$ has been implemented.
It is because of the $\numu$--$\nutau$ interchange symmetry
that the final result depends on the single weight $\we$.
This greatly reduces the latitude in characterizing the flavor
distribution at the source.

For neutrinos with energies well below any matter resonances,
the matter effects are negligible.
These neutrinos therefore propagate as if in vacuum,
and for tribimaximal mixing they produce flavor probabilities at Earth 
given by the formulas above.  
At higher energies, 
matter effects become substantial.
One test for the efficacy of the solar matter potential on neutrino mixing 
would be to determine flavor probabilities with values
{\sl other} than those 
presented in Eqs.~(\ref{probvacuum2a}) and (\ref{probvacuum2b}).

\subsection{Oscillations with matter: \\
   adiabatic approximation}
\label{subsec:matteroscillations}
In matter, the forward scattering amplitude of the neutrino 
leads to an effective potential~\cite{MSW}.  
The scattering amplitudes common to all 
neutrino flavors generate a common potential, which is unobservable in 
neutrino oscillations.  However, there is a scattering amplitude unique to
$\nue$ and $\nuebar$ flavors.  It is the $W$-exchange diagram for scattering 
on background electrons, with exchanges in the $t$-channel for $\nue$ and 
in the $s$-channel for $\nuebar$.  Thus, there results 
an effective potential for the $\nu_e$ and $\nuebar$ flavor states,
given by $V_e(r)=\sqrt{2}\,G_F\,n_e(r)$ for $\nu_e$, and $-V_e$ for $\nuebar$, 
where $n_e$ is the electron number density.
The effect of the matter contribution on neutrino flavor mixing is best 
found by first transforming the vacuum Hamiltonian~\rf{Hmassvacuum} to the 
flavor basis, labeled by $\{|e\ra,|\mu\ra,|\tau\ra\}$,
and then adding to this the flavor-diagonal effective potential.
Then, 
the flavor-basis effective Hamiltonian in matter
\beq{eff_ham}
{}\hspace{-1.8mm}H_{F}=
U\,\frac{{\rm diag}(-\delta m_{21}^2,0,\delta m_{32}^2)}{2E}\,U^{\dagger}
+V_e\,{\rm diag}(1,0,0)
\eeq
emerges.
In Eq.~(\ref{eff_ham}), 
some contributions to $H_F$  proportional to the identity have been omitted for convenience 
because they leave unchanged the flavor mixing, 
as explained earlier.
For antineutrinos,
the sign of matter-potential term is reversed
and $U$ is replaced by $U^*$.

The matrix 
that diagonalizes the flavor Hamiltonian in the 
matter background is called $U_m$:
$H_M=U_m^\dagger\,H_F\,U_m$. 
Such a matrix is only fixed modulo the ordering and the phases of its column vectors. 
The phases are without physical significance 
because they can be absorbed into the definition of the states $|j,r\ra$ 
defined below. 
The ordering of the column vectors, 
on the other hand, 
determines the order of the eigenvalues along the diagonal of $H_M$. 
Our convention is this: 
the order of the column vectors in $U_m$ is such 
that the diagonal entries of $H_M$ 
exhibit the same ordering in magnitude 
as those of the matrix $\hat{M}={\rm diag}(-\delta m_{21}^2,0,\delta m_{32}^2)$ 
in Eq.\ \rf{eff_ham}. 
Note that this convention depends on the mass hierarchy,
via the sign of $\delta m_{32}^2$.
This convention is chosen 
such that $U_m$ can smoothly approach the vacuum mixing matrix $U$ 
in the limit of vanishing matter potential. 

Given a position-dependent matter potential, 
the mixing matrix $U_m (r)$ will also depend on position.
For the matrix which diagonalizes $H_F$ {\sl at the Sun's center},
we reserve the notation $U_m$ without argument $r$.
In the solar core, the matter density plateaus, 
and the matter potential at the Sun's center is 
\beq{V(0)}
V_e (0)=70\,\sqrt{2}\,G_F\;N_A\,{\rm cm}^{-3}=5.6\times 10^{-12}\,{\rm eV}\,.
\eeq
Thus, neutrinos from the solar core with energy above 
$\sim \delta m^2_{jk}/2V_e(0)\sim (\delta m^2_{jk}/10^{-5}{\rm eV}^2)\,{\rm MeV}$
will feel the effect of the Sun's matter potential.
This class includes the $E\gsim\,$GeV neutrinos from WIMP annihilation.

With the mixing matrix in matter $U_m (r)$
and the flavor eigenstates at hand, 
one can construct the ``instantaneous'' eigenstates of $H_F$
via $|j,r\ra=U_m^\dag (r) |\alpha\ra$. 
Although these states
are strictly speaking not solutions to the equations of motion, 
they will turn out to be invaluable in what follows. 
For future reference,
we further define the set of ``matter states'' $\{|1\ra_m,|2\ra_m,|3\ra_m\}$
as the exact solution 
of the Schr\"odinger Equation with the Hamiltonian $H_F$
given in Eq.\ \rf{eff_ham}.
It is typically difficult 
to determine explicit expressions for these states.
In this section, 
we simply note 
that equating $|j\ra_m$ to $|j,r\ra$
is known as the ``adiabatic approximation.''
We defer to Sec.~\ref{valid} a more detailed discussion of this approximation.
To summarize,
the weak-interaction Hamiltonian
is diagonal in the flavor basis $|\alpha\ra$. 
On the other hand, 
the free-neutrino Hamiltonian in vacuum or matter is diagonal
in the mass basis $|j\ra$ or the matter basis $|j\ra_m$, 
respectively.
An approximation for the matter states 
is given by the instantaneous eigenstates $|j,r\ra$.

\subsubsection{Diagonalization of $H_F$ for $\theta_{13}=0$}
\label{subsec:diagonalizing}
The diagonalization of $H_F$ is simplified
by the observational input 
that $\theta_{13}$ is small,
so we can initially approximate it by zero.
Then, 
the $3\times 3$ vacuum mixing matrix simplifies to
$U=R_{23}(\theta_{23})\,R_{12}(\theta_{12})$.
This permits us to cast $H_F$ into block-diagonal form
via the transformation 
$H'_F\equiv R^{-1}_{23}(\theta_{23})\,H_F\,R_{23}(\theta_{23})$
because the rotation matrix $R_{23}$ commutes with the matter-potential
term in Eq.\ \rf{eff_ham}.  The result is 
\beq{part_diag1}
H'_F=
\left(\begin{array}{cc}
H_{2\times2} & \begin{array}{c}0\\0\end{array}\\
\begin{array}{cc}0&0\end{array}&\displaystyle{\frac{\delta m_{32}^2}{2E}}
\end{array}\right)\,,
\eeq
with
\beq{part_diag2}
{}\hspace{-1mm}H_{2\times2}=
\left(\!\!\begin{array}{cc}
V_e(r)-\displaystyle{\frac{\delta
m_{21}^2}{2E}}\cos^2\theta_{12}\vspace{1mm} &
\displaystyle{\frac{\delta m_{21}^2}{4E}}\sin2\theta_{12}\vspace{1mm}\\
\displaystyle{\frac{\delta m_{21}^2}{4E}}\sin2\theta_{12} &
-\displaystyle{\frac{\delta m_{21}^2}{2E}} \sin^2\theta_{12}\\
\end{array}\!\right)
\eeq
determining the upper $2\times2$ block.
Note that the matter potential $V_e(r)$
and off-diagonal elements are confined to $H_{2\times2}$.
This means that the $|3\ra_m$ mass eigenstate in matter is
unaffected by background electrons 
and decouples from the other two states.
This allows a two-state treatment
of the original three-flavor problem
leading to some simplifications,
which we discuss next.

Since the state $|3\ra_m$ is unaffected by the background matter,
it is identical to the vacuum mass eigenstate
$|3\ra=(R^{-1}_{23})_{3\alpha}\,|\nu_\alpha\ra
=\sin\theta_{23}|\mu\ra+\cos\theta_{23}|\tau\ra$.
The fact that $|3\ra_m=|3\ra$ decouples from the other two states
implies that the currently unknown neutrino mass hierarchy
($m_3>m_2>m_1$ vs.\ $m_2>m_1>m_3$)
is unimportant in the present case characterized by $\theta_{13}=0$.

Another simplification following from the three- to two-state reduction
is the elimination of any possible CP-violation 
from the effective mixing matrix.
This is evident 
in that the phase $\delta$ 
does not enter into Eqs.~(\ref{part_diag1}) and (\ref{part_diag2}).
Still, 
the neutrino and antineutrino cases have to be treated separately,
for their respective Hamiltonians differ
in the sign of the matter-potential term.

A third simplifying feature 
arising from the reduction to two-state mixing 
is the straightforward recognition of a resonance condition.
Mixing becomes resonant 
(maximal angle and degenerate eigenvalues) 
when the two diagonal elements of $H_{2\times 2}$ become equal,
i.e., when 
\beq{Eres}
E_R(r)=\delta m^2_{21}\,\cos2\theta_{12}/2\,V_e(r)\,.
\eeq
The resonant energy at the Sun's center is therefore
\beq{Erescenter}
E_R (0)=\cos 2\theta_{12}\,(\delta m^2_{21}/10^{-4}{\rm eV}^2)\,10\,{\rm MeV}\,.
\eeq
The tribimaximal value for $\cos 2\theta_{12}$ is 1/3.
Equation~\rf{number_density} implies 
that outside the solar core 
the matter potential decreases, 
so that the resonance energy increases.
Thus,
only neutrinos with energies
exceeding the solar-core value \rf{Erescenter}
can experience resonant conversion.
The sign of $\delta m^2_{21}$ is fixed as positive by solar data.
It follows 
that for antineutrinos the right-hand side of Eq.~\rf{Eres} is negative,
and there is no possibility of resonance conversion when $\theta_{13}=0$.

The next step is to determine the eigenstates $|1\ra_m$ and $|2\ra_m$
of the effective $2\times 2$ Hamiltonian \rf{part_diag2}.
Following the conventional analysis~\cite{review},
we approximate these states
by the instantaneous eigenvectors $|1,r\ra$ and $|2,r\ra$
that diagonalize $H_{2\times 2}$ at the point $r$.
This adiabatic approximation is valid 
if the matter density changes slowly
enough with distance.
In this section, 
we assume adiabaticity holds.
In the next section, 
we relax this assumption.

In the limit 
\beq{low_limit}
E\gg \delta m^2/2\,V_e(0)\;, 
\eeq
the instantaneous eigenstate 
$|1,0\ra$ at the Sun's core $r\simeq 0$ is predominantly determined by the matter potential,
leading to $|1,0\ra\simeq|e\ra$.
The majority of neutrinos from solar WIMP annihilation
are expected to have energies $E\gsim{\cal O}({\rm GeV})$;
for these neutrinos, $|1,0\ra\simeq|e\ra$ is a good approximation~\cite{Umremark}.
Then the only mixing is due to $\theta_{23}$ and we have 
$U_m=R_{23}(\theta_{23})$~\cite{Umremark}.
Thus, 
the remaining instantaneous mass state
is given by 
$|2,0\ra\simeq\cos\theta_{23}|\mu\ra -\sin\theta_{23}|\tau\ra$.

As for the vacuum case in the previous subsection,
we employ a density-matrix analysis
to predict the flavor evolution. 
We again begin 
with the density matrix $\rho(0)$ 
describing the mixed-flavor ensemble
produced at the solar core.
Our above considerations 
regarding the instantaneous mass states at the Sun's center yield~\cite{Umremark}
\bea{dmxprodn}
\rho(0) & = &w_e\;|e\ra\la e|+w_\mu\;|\mu\ra\la \mu|+w_\tau\;|\tau\ra\la \tau|\nonumber\\
        & = &w_e\,|1,0\ra\la 1,0|
    + w_\mu\left( \frac{|3\ra +|2,0\ra}{\sqrt{2}}\right)\!
        \left( \frac{\la3| +\la 2,0|}{\sqrt{2}}\right)\nonumber\\
    && {}+w_\tau\left( \frac{|3\ra -|2,0\ra}{\sqrt{2}}\right)\!
        \left( \frac{\la3| -\la 2,0|}{\sqrt{2}}\right).
\eea
In the latter expression, 
the best-fit value of $\theta_{23}\simeq 45^\circ$
has been implemented.

We continue by 
studying the propagation of the neutrinos from 
the Sun's core to its surface.
Quantum interference effects can be safely neglected 
because the neutrinos quickly decohere after production:
With the exception of possible resonance points $r_{\rm res}$ 
where the two eigenvalues of $H_{2\times2}$ approach each other,
the oscillation lengths $\lambda$ are small compared to solar scales.
For example,
in the production region $r\simeq 0$,
the set of approximate instantaneous eigenvalues of $H'_F$ 
is $\{-\delta m_{12}^2/2E, \delta m_{32}^2/2E, V_e(0)\}$.
The relative sizes 
of $\delta m_{32}/2E$ and $V_e(0)$ dominate,
so that the oscillation scale is given by 
$E/\delta m_{32}^2\lsim{\cal O}(100\,{\rm km})$ 
for $E>1\,$GeV, $\delta m_{32}^2\approx 10^{-3}\,$eV$^2$, 
and $V^{-1}_e(0)\sim 40\,{\rm km}$.
If the uncertainty in $\delta L$ is comparable to these oscillation lengths,
then the phase of the oscillation is randomized per neutrino 
and averaged out in the statistical sense.
Such is the case here, 
since the size $\sim 0.01R_\odot\sqrt{100\;\textrm{GeV}/m_\textrm{DM}}$ 
of the WIMP annihilation region in the Sun's core~\cite{dm_trap} 
exceeds the oscillation scale of 100~km.
Due to this decoherence,
we can transform the density matrix $\rho$ to the mass basis,
and then disregard the off-diagonal pieces,
as was done for the vacuum case.
The result is three incoherent fluxes associated with the matter 
states $|1,0\ra$, $|2,0\ra$, and $|3\ra$
emerging from the production region at the center of the Sun.

Neutrinos corresponding to $|3\ra$
will remain in this state
because it is an exact eigenstate of the Hamiltonian
unaffected by the solar electron density.
However,
the instantaneous eigenstates $|1,r\ra$ and $|2,r\ra$
change with respect to the flavor basis
as $r$ increases,
and transitions between these states can occur.
The adiabatic approximation
neglects these transitions,
so that a particle will remain 
in an instantaneous eigenstate.
We note
that the flavor-oscillation wavelengths for GeV neutrinos 
close to resonances in the Sun 
are long compared to the $0.1\ R_{\odot}$ scale length 
of the exponentially decreasing solar density,
so that the adiabatic approximation breaks down
in some circumstances.
The adiabatic analysis presented here 
is nevertheless interesting, 
for it is instructive, simple, and applicable in certain energy ranges,
as we shall see.

We are now in the position
to establish the evolution of the density matrix.
The discussion above shows
that at sufficiently large neutrino energies
the vacuum masses are negligible compared
to the effective mass of the state $|1,0\ra$.
However,
outside the Sun,
where $V_e(r>R_\odot)= 0$,
the mass hierarchy is reversed:
the instantaneous states
are just the vacuum mass states
$|1,r>R_\odot\ra=|1\ra$ and $|2,r>R_\odot\ra=|2\ra$
characterized by $m_1<m_2$.
Since the usual quantum-mechanical level repulsion
prohibits level crossing,
the states in the adiabatic approximation evolve as follows:
as the neutrino travels to the solar surface,
we have $|1,0\ra\rightarrow|2\ra$
and $|2,0\ra\rightarrow|1\ra$
up to phases~\cite{Umremark}.
As a result, 
we obtain the density matrix
\beq{dmxearth}
{}\hspace{-1mm}\rho(r>R_\odot) = w_e\,|2\ra\la 2|
    + \half\,(w_\mu+w_\tau)\,(|3\ra\la 3| +|1\ra\la1|)
\eeq
for neutrinos emerging from the Sun.
Upon exiting the Sun,
the $|j\ra$ states become exact eigenvectors of the neutrino Hamiltonian,
so that transitions between these states
cease to occur.
Outside the Sun in empty space, 
the density matrix therefore remains effectively unchanged 
obeying $\rho ({\rm Earth})=\rho(r>R_\odot)$.

This result can be employed
to determine the relative flavor fluxes
measured in Earth-based experiments. 
The probability of detecting the neutrino flavor $\beta\in\{e,\mu,\tau\}$
by such experiments is
\bea{probalpha}
{}\hspace{-6mm}P_{\nu_\odot\rarr\nu_\beta} 
& \!=\! & \la\beta|\rho({\rm Earth})|\beta\ra\nonumber \\
& \!=\! & w_e\,|U_{\beta 2}|^2
        +\half\,(w_\mu+w_\tau)\left(1-|U_{\beta 2}|^2\right)\,.
\eea
In the final step,
we have used $\sum_{j=1}^3 |U_{\beta j}|^2 =1$,
which arises from the unitarity of $U$.
The $\numu$--$\nutau$ interchange symmetry is evident again.
The relation $\we+\wmu+\wtau=1$
allows us to cast the expression in terms of $\we$ only.
We remind the reader 
that although this result incorporates solar matter effects,
the $|U_{\beta j}|^2$ in the final expression 
are elements of the {\sl vacuum} matrix $\U2$.

Note that an amusing result emerges if 
$|U_{e2}|^2=|U_{\mu 2}|^2=|U_{\tau 2}|^2$,
namely, that all three flavors arrive at Earth with equal probability 1/3,
{\sl independent of the flavor ratios at production}.
While $|U_{e2}|^2=|U_{\mu 2}|^2=|U_{\tau 2}|^2$ may seem contrived,
in fact it is the cornerstone of the tribiximal 
mixing matrix.  We discuss this next.

\subsubsection{Neutrino flavor evolution}
\label{subsec:nu-evo}
To obtain the expected neutrino flavor probabilities at Earth, 
we insert the tribimaximal mixing values for $\U2$
given in Eq.~\rf{tribimaxsq}, into Eq.~(\ref{probalpha}).
Doing so, 
and using the normalization relation $\sum_\alpha w_\alpha =1$,
we arrive at
\beq{probs2}
P_{\nu_\odot\rarr\nu_e}=P_{\nu_\odot\rarr\nu_\mu}=P_{\nu_\odot\rarr\nu_\tau} 
  =  \frac{1}{3}\,.
\eeq
As advertised above, this result for neutrinos 
is independent of the $w_\alpha$
which characterize the flavor distribution at the Sun's center.
Such a result could hardly be simpler (or more democratic).
It is easily understood as well:
With the assumption that $\theta_{13}=0$, 
the state $|3\ra$ is completely decoupled, 
and so the higher-energy resonance is absent.
Thus, a $\nu_e$ produced with energy above the lower-energy resonance 
emerges from the Sun as pure $|\nu_2\ra$ in the assumed adiabatic limit.
With tribimaximal mixing,
the decomposition of $|\nu_2\ra$ into flavors gives the observed 
ratios $|U_{e2}|^2 : |U_{\mu 2}|^2 : |U_{\tau 2}|^2 = 1:1:1$.
On the other hand, 
$\nu_\mu$'s and $\nu_\tau$'s produced in the Sun's center
equilibrate with each other, 
a consequence of the $\nu_\mu$--$\nu_\tau$ interchange symmetry.
They then emerge from the Sun with equal parts of the two states orthogonal to 
$\nu_2$, i.e., with equal parts of $\nu_1$ and $\nu_3$.
The flavor decomposition of these states is 
$|U_{\beta 1}|^2 +|U_{\beta 3}|^2 = 1-|U_{\beta 2}|^2$,
which is again $1:1:1$ for the tribimaximal mixing matrix.
Thus, $\nu_e$ and independently $\nu_\mu$ and $\nu_\tau$ generate a 
$1:1:1$ flavor ratio at Earth.
As a consequence, the $1:1:1$ prediction is independent of the initial $w_\alpha$.
The essence of this argument is visible in Eqs.~(\ref{dmxearth}) and (\ref{probalpha}).

Of course, the state evolution is not perfectly adiabatic,
and $\theta_{13}$ is probably not exactly zero.
With either alteration, one expects deviation from the $1:1:1$ flavor ratios.
We address these issues in Sec.~\ref{sec:non-adiabatic}.
Figure~\ref{fig1} shows a comparison of our simplified approach
described here with the full non-adiabatic result obtained in Sec.~\ref{sec:non-adiabatic}.

We remark that the democracy derived above for the arriving neutrino flavors
is a nontrivial prediction.  First of all, it depends on the approximate validity of the 
tribimaximal mixing matrix.
Second, it depends on the solar matter, as outlined above.  
Reference to the vacuum predictions 
given in Eqs.~\rf{probvacuum2a} and \rf{probvacuum2b} 
shows that without the solar-matter effect, 
flavor democracy results only for the single value $\omega_e=1/3$.  
Third, even with tribimaximal mixing and the solar-matter effect,
democracy is not a prediction for the arriving antineutrino flavors,
as we show next.

\subsubsection{Antineutrino flavor evolution}
Antineutrinos from the Sun yield more interesting flavor probabilities.
The construction of these flavor probabilities
follows the same logic as those for neutrinos.
However, there is one major difference:
the hierarchy of the effective mass in the solar core
and the vacuum mass hierarchy are identical,
so that no mass-label permutation is necessary.
To obtain the antineutrino analogue of Eq.\ \rf{dmxearth},
we must therefore take $|1,0\ra\rightarrow|1\ra$
and $|2,0\ra\rightarrow|2\ra$ in Eq.\ \rf{dmxprodn}.
This gives
\beq{probalphabar}
P_{\nubar_\odot\rarr\nubar_\beta} = \webar\,|U_{\beta 1}|^2
        +\half\,(\wmubar+\wtaubar)(1-|U_{\beta 1}|^2)\,.
\eeq
The tribimaximal mixing probabilities are 
$|U_{e1}|^2=\frac{2}{3}$ 
and $|U_{\mu 1}|^2 =|U_{\tau 1}|^2 =\frac{1}{6}$.
It follows that
\bea{probanti}
P_{\nubar_\odot\rarr\nubar_e}&=&\frac{1}{6}\;(1+3\,\webar)\,,\nonumber\\
P_{\nubar_\odot\rarr\nubar_\mu}&=&P_{\nubar_\odot\rarr\nubar_\tau}
    =\frac{1}{12}\;(5-3\,\webar)\,.
\eea
Here, the answer does depend on the flavor ratios $w_{\bar \alpha}$
at production in the Sun's center.
For example, an injection flux of purely $\nuebar$ leads to 
$P_{\nubar_\odot\rarr\nubar_e}=\frac{2}{3}$ and
$P_{\nubar_\odot\rarr\nubar_\mu}=P_{\nubar_\odot\rarr\nubar_\tau}
  =\frac{1}{6}$,
whereas an injection flux with no $\nuebar$ leads to 
$P_{\nubar_\odot\rarr\nubar_e}=\frac{1}{6}$ and
$P_{\nubar_\odot\rarr\nubar_\mu}=P_{\nubar_\odot\rarr\nubar_\tau}=\frac{5}{12}$.
Thus, in principle the $w_{\bar \alpha}$ are measurable.  
Their determination would help unravel the nature of the DM source.

Why are the antineutrino flavor ratios not democratic, as were the neutrino 
ratios?
The answer lies in the fact that $\nuebar$ emerges as pure ${\bar\nu}_1$,
which is distributed in flavor as $4:1:1$,
and the $\numubar$ and $\nutaubar$ emerge as equal parts 
${\bar\nu}_2$ and ${\bar\nu}_3$,
which are collectively distributed as $2:5:5$.
Thus, the resulting flavor ratios depend on the initial ratios $w_{\bar \alpha}$.

Since resonances in the antineutrino sector are absent within the normal hierarchy,
another question arises.
Why do the antineutrino results here differ from the vacuum 
results of Section \ref{subsec:vacuumoscillations}?
The difference is due to the mixing effect at the antineutrino production site.
The large matter potential causes $|e\ra$ to be almost purely $|1\ra_m$,
whereas in vacuum $|e\ra$ would be distributed among the mass states $|j\ra$ 
as $|U_{ej}|^2$.

In Fig.~\ref{fig1}, 
the approximate antineutrino probabilities derived here for 
adiabatic evolution and $\theta_{13}=0$
can be compared to the full analysis performed in the next section.
It is seen in Fig.~\ref{fig1} that the simple approximation
characterized by constant (i.e., energy-independent) probabilities, 
is good over a relevant range of several decades in energy for the antineutrino.

We end this section with the remark 
that in both the neutrino and the antineutrino cases,
all flavor probabilities converge to 1/3 (democratic $1:1:1$) 
for $w_e$ and $\webar=1/3$.
This is because for these latter values all
mass eigenstates are equally represented at production,
and subsequent level repulsion simply permutes the labels of these
mass states.
This result holds for any values of the mixing matrix,
and it also applies in the presence of matter effects, 
as we prove in Sec.~\ref{subsec:nonadiabatic}.

\subsubsection{Validity range of the above simplified results}
\label{valid}
As is evident in Fig.~\ref{fig1}, for the normal mass hierarchy,
the adiabatic approximation is good for neutrinos with energies 
between $\sim10$~MeV and $\sim10$~GeV,
and for antineutrinos with energies above $\sim10$~MeV.
What physics sets these limits of validity?

The low-energy limit of validity $E_{\rm low}$ is determined by 
the requirement that at production, the matter state $|e\rangle$ be well
separated from the other states; in turn, this requires that  
the matter-potential term dominates the vacuum mass term
in the effective Hamiltonian \rf{eff_ham} at production.
A sufficient condition for this stipulation is given by Eq.~\rf{low_limit},
which yields $E_{\rm low}\simeq 70$ MeV when the value of $\delta m^2_{32}$
is taken.  
In fact, 
the limit $E_{\rm low}$ can even be lower
due to the particular form of the mixing matrices 
in the Hamiltonian~\rf{eff_ham}.

At energies well below the lowest resonance,
the matter term in the Hamiltonian becomes negligible and 
vacuum results are approached.  As can be seen in Fig.~\ref{fig1},
the low-energy asymptotes for all neutrino and antineutrino curves have the vacuum values
given in Eqs.~\rf{probvacuum2a} and~\rf{probvacuum2b}.

The high-energy limit of validity $E_{\rm high}$
is determined by the onset of non-adiabaticity,
i.e., when appreciable transitions between the instantaneous eigenstates begin to occur.
The non-adiabaticity scale $\TEna$ is established in the next section.
Borrowing results from the next section, we can conclude the following:\ 
$\TEna\rightarrow\infty$ for the nonresonant $\bar{\nu}$ case,
and from Eq.\ \rf{Ena}, $\TEna\sim 100$~GeV 
for the resonant $\nu$ case with the parameters listed in Fig.~\ref{fig1}.

Neutrino absorption in the Sun is significant at energies above $\sim 100$~GeV, 
thus providing another limit on the validity of our analysis.
If the absorption is universal across the flavors, 
then one might expect 
a reduction in flux and a change in spectrum,
but little or no alteration of the flavor results we 
derive in this paper.

\subsubsection{Neutrinos and antineutrinos together}
\label{together}

An unmagnetized detector is ill-suited to resolve $\nu$ events from $\nubar$ events.
For such a detector it is more appropriate
to consider the weighted average
\beq{nuplusnubar}
\la P_{\nu_\odot\rarr\nu_\beta\,{\rm or}\,\nubar_\beta}\ra = 
\frac
{\sigma_\nu\,F_{\nu_\odot}\,P_{\nu_\odot\rarr\nu_\beta}+
\sigma_{\nubar}\,F_{\nubar_\odot}\,P_{\nubar_\odot\rarr\nubar_\beta}}
{\sigma_\nu\,F_{\nu_\odot}+\sigma_{\nubar}\,F_{\nubar_\odot}}
\eeq
as an observable. 
Here, $F_{\nu_\odot}$ and $F_{\nubar_\odot}$ 
are the neutrino and antineutrino fluxes produced by WIMP annihilation in the Sun,
and $\sigma_{\nu}$ and $\sigma_{\nubar}$ are the neutrino and antineutrino 
cross sections in the detector.
We define the neutrino to antineutrino cross-section ratio as 
$r_\sigma\equiv \sigma_{\nubar}/\sigma_\nu$.
Experimentally, $r_\sigma \sim 1/2$ in the energy range of interest.
In situations in which neutrino production is charge symmetric, the initial 
$\nu_\alpha$ and $\nubar_\alpha$ fluxes are equal, as are
$w_\alpha$ and $\walphabar$.
The weighted average of Eqs.~\rf{probalpha} and \rf{probalphabar} 
is then simply given by
\begin{widetext}
\beq{probsym}
\la P_{\nu_\odot\rarr\nu_\beta\,{\rm or}\,\nubar_\beta}\ra =
\frac{
  \left[
\we\,(3\,|U_{\beta 2}|^2-1)+1-|U_{\beta 2}|^2
\right] \\
+\;r_\sigma\,
\left[
\we\,(3\,|U_{\beta 1}|^2-1)+1-|U_{\beta 1}|^2
\right] 
}
{2\,(1+r_\sigma)}
\eeq
\end{widetext}
Assuming the tribimaximal mixing angles, one obtains
\bea{probsym2}
\!\!\!\!\!\la P_{\nu_\odot\rarr\nu_\mu\,{\rm or}\,\nubar_\mu}\ra &=&
\la P_{\nu_\odot\rarr\nu_\tau\,{\rm or}\,\nubar_\tau}\ra
=\frac{4+r_\sigma(5-3\we)}{12\,(1+r_\sigma)}\,,\nonumber\\
\!\!\!\!\!\la P_{\nu_\odot\rarr\nu_e\,{\rm or}\,\nubar_e}\ra &=&
\frac{2+r_\sigma(3\we+1)}{6\,(1+r_\sigma)} \,.
\eea
The $w_e$ dependence of these results, although  weakened compared to 
the pure $\nubar$ sample 
given in Eq.~(\ref{probanti}),
implies that detectors that sum $\nu$ and $\nubar$ events 
still have some discriminatory power
to resolve different source models.
In Sec.~\ref{sec:models} we discuss this possibility.

\subsubsection{Summary of adiabatic evolution in matter}
\label{subsec:summary}
In our description of adiabatic evolution in matter, 
we adopted three simplifying assumptions governing 
neutrino (and antineutrino) production by solar 
WIMP annihilation, and subsequent propagation out of the Sun to Earth.
These assumptions are
that all neutrino energies exceed the resonant energy at the Sun's center 
($E_R (0)\sim 10$~MeV),
that the neutrinos evolve adiabatically through the Sun,
and that neutrino mixing angles are given by the tribimaximal values,
including $\theta_{13}=0$.
As a consequence, we obtained energy-independent flavor probabilities for 
the neutrinos arriving at Earth.
We learned that solar matter plays a significant role in 
determining the expected flavor ratios at Earth.
However, the higher-energy resonant value $E_R^h$ has yet to play a role,
since $\theta_{13}=0$ decouples the $|3\ra$ state from the $|1\ra$
and $|2\ra$ states.
In the next subsection, we relax the various assumptions.

\subsection{Non-adiabatic oscillations and nonzero $\bm\theta_{13}$}
\label{sec:non-adiabatic}
The physics becomes more complicated when simplifying assumptions are abandoned.
For example,
if $\theta_{13}$ is no longer assumed to be zero, 
then the state $|3\ra$ no longer decouples 
and a full three-flavor analysis with two resonant energies is typically required.

\subsubsection{Nonzero $\theta_{13}$ and three-flavor mixing}
\label{subsec:threeflavor}
We begin again
with the density matrix $\rho_{\nu}(0)$ at production,
expressed in flavor space:
\beq{dmx0}
\rho_{\nu}(0)
=\sum_\alpha w_\alpha\;|\alpha\ra\la\alpha|\,,
\eeq
where $\alpha\in\{e,\mu,\tau\}$ runs over the three flavors,
and the $w_\alpha$ are the relative flavor fluxes at production,
as before.
Paralleling our previous simplified analysis,
we expand the flavor eigenstates $|\alpha\ra$
in the basis of instantaneous eigenstates $|j,r\ra$ at $r=0$, 
where $j\in\{1,2,3\}$ labels again the effective-mass eigenvalue:
$|\alpha\ra=\sum_j \la j,0\,|\,\alpha\ra\;|j,0\ra$.
We obtain
\beq{dmx1}
\rho_{\nu}(0)
=\sum_{\alpha,j,k} w_{\alpha} \,
(U_m)_{\alpha j}\,(U_m)^*_{\alpha k}\,|j,0\ra\la k,0|\,.
\eeq
Recall the earlier definition of $U_m$ as the solar-core
instantaneous mixing matrix in matter, 
i.e., $(U_m)_{\alpha k}\equiv\la\alpha\,|\,k,0\ra$.
The matter at the Sun's center characterizes the core 
environment where the neutrinos are produced.
Again,
as so defined,
$U_m$ diagonalizes $H_F$ only at the center of the Sun.

The same decoherence arguments presented in the previous section
allow us again to neglect the off-diagonal density-matrix elements in Eq.\ \rf{dmx1}.
Therefore, in what follows we will take
\beq{dmx2}
\rho_{\nu}(0)
=\sum_{\alpha,j} w_{\alpha}\,|(U_m)_{\alpha j}|^2\,|j,0\ra\la j,0|
\eeq
as the neutrino-production density-matrix.
As a check, 
we note that
with a large matter potential isolating the $|\nu_e\ra$ state 
and maximal $\theta_{32}=45^\circ$, 
Eq.\ \rf{dmx2} becomes identical to our earlier Eq.~(\ref{dmxprodn}).

\subsubsection{Non-adiabatic evolution}
\label{subsec:nonadiabatic}
Next,
we re-examine the evolution of this system
during propagation to the solar surface,
but this time without the adiabatic assumption of the previous section.
That is, we now allow for transitions
between the instantaneous eigenstates $|j,r\ra$.

Once the neutrinos reach the vacuum at the Sun's surface,
transitions between the eigenstates 
$|j,r>R_\odot\ra=|j\ra$ no longer occur
because in free space the $|j\ra$ are exact eigenstates of the Hamiltonian.
For our analysis,
we are therefore interested in the solar-propagation amplitudes
$A_{kj}\equiv\la k|j,0\ra$.
Embedded in these amplitudes are the important non-adiabatic transitions between 
instantaneous eigenstates, most likely to occur near resonance.

In terms of the amplitudes $A_{kj}$, the density matrix given in Eq.~\rf{dmx2} 
evolves to become, beyond the Sun's surface,
\beq{dmx3}
\rho_{\nu}(r>R_\odot)
=\sum_{\alpha,j,k,l} w_{\alpha}\;
|(U_m)_{\alpha j}|^2 \,A^*_{lj}A_{kj}\,|k\ra\la l|\,.
\eeq
It is seen here that non-adiabatic transitions replenish the 
off-diagonal elements of $\rho$.
However, the $A_{kj}$ in general contain position-dependent phases,
which must be averaged over the production region.
Furthermore, additional loss of phase information
occurs as the neutrinos continue en route to Earth,
as discussed in Sec.~\ref{subsec:vacuumoscillations}.
So again the off-diagonal terms average to zero.  
Consequently, Eq.~\rf{dmx3} is reduced to the more manageable expression
\beq{dmx4}
\rho_{\nu}({\rm Earth})
=\sum_{\alpha,j,k} w_{\alpha}\; 
|(U_m)_{\alpha j}^{}|^2 \,|A_{kj}|^2\, |k\ra\la k|\,.
\eeq
The amplitude $A_{jk}$ is seen to enter the final formula 
only via the jump probability $P_{kj}\equiv |A_{kj}|^2$,
as expected in the absence of coherence.

Equation \rf{dmx4} can now be employed 
to determine the probability 
of detecting a neutrino of flavor $\beta$ at Earth. 
We obtain 
%
\bea{fullprobalpha}
P_{\nu_\odot\rarr\nu_\beta}&=& \la\beta|\rho({\rm Earth})|\beta\ra\nonumber\\
&=&\sum_{\alpha,j,k} w_{\alpha} \;
|(U_m)_{\alpha j}|^2 \,P_{kj}\,|U_{\beta k}|^2\,.
\eea
In matrix form, this equation is given by
\beq{matrixform2}
\left(
\ba{l}
P_{\nu_\odot\rarr\nu_e}\\
P_{\nu_\odot\rarr\numu}\\
P_{\nu_\odot\rarr\nutau}
\ea
\right)
 = \U2\,P\,\U2_{\hspace{-.3mm}m}^T
\left(
\ba{l}
\we\\
\wmu\\
\wtau
\ea
\right)\,.
\eeq
Here, we have defined the matter matrix 
$(\U2_{\hspace{-.3mm}m})_{\alpha j}\equiv |(U_m)_{\alpha j}|^2$,
in analogy to our earlier vacuum definition
Equations \rf{fullprobalpha} and \rf{matrixform2}
generalize Eqs.~\rf{probalpha} and \rf{matrixform} 
in the previous subsections.
In the adiabatic ($P_{kj}\rarr\delta_{kj}$) and matter-free ($\U2_{\hspace{-.3mm}m}\rarr \U2$) limits,
these equations reduce to their analogues in the previous subsections. 

With Eq.~\rf{fullprobalpha} at hand, 
we may establish rigorously an interesting feature of the flavor evolution:
{\sl If the flavors at production are democratically distributed
($\we=\wmu=\wtau=1/3$), 
then the flavor probabilities observed at Earth will also 
be democratically distributed.}
In other words,
the particular ratio $(\frac{1}{3},\frac{1}{3},\frac{1}{3})$ 
of flavors is left unchanged
by the combination of all oscillation and matter effects.
This is expected in that off-diagonal density-matrix elements average to zero, 
so that the density-matrix evolves by changing bases, from flavor in matter 
to evolving mass in matter, to constant mass in vacuum, and finally to flavor at earth;
but a density matrix proportional to the unit matrix initially, remains so throughout 
basis changes.

We verify this conclusion 
by setting $\walpha=1/3$ in Eq.~\rf{fullprobalpha},
and rewriting the remaining terms slightly:
\beq{spec_prob}
P_{\nu_\odot\rarr\nu_\beta} 
=\frac{1}{3}\,\sum_{k}|U_{\beta k}|^2\,\sum_{j}P_{kj}\,
   \sum_{\alpha}|(U_m)_{\alpha j}|^2\,.
\eeq
The unitarity of $U_m$ implies that the sum over $\alpha$
appearing on the far right-hand side of Eq.\ \rf{spec_prob}
equals unity.
Similarly, the sum of elements in the row or column of $P_{kj}$ 
must add up to 1 because of probability conservation.
The remaining sum over $k$
is also equal to 1 due to the unitarity of $U$.
We are thus left with the claimed result
that $P_{\nu_\odot\rarr\nu_\beta} 
=\frac{1}{3}$ for all flavors $\beta$.
Consequently, any experimentally inferred flavor ratios deviating from 
$1:1:1$ exclude a democratic source.

Equation \rf{fullprobalpha} allows predictions to be made for neutrino flavors at Earth,
given an initial distribution of flavors in the solar core.
It has been used extensively in the context of suppression of 
the expected solar $\nu_e$ rate at Earth.  For neutrinos from the solar fusion cycle,
one has $w_e=1$, with all other $w_\alpha$ and $w_{\bar\alpha}$ equal to zero.
However, little study of Eq.~\rf{fullprobalpha} has been performed for an ensemble of 
solar flavors, as would arise from WIMP annihilation in the Sun.
An early study of neutrino flavor physics arising from solar WIMP annihilation 
was performed in Ref.~\cite{ellis92}.
Much has been learned about the neutrino mixing matrix in the intervening 
fifteen years. We incorporate subsequent knowledge into our work here.

An approximation for $\U2_{\hspace{-.3mm}m}$ is derived in Appendix~\ref{matter_mixing},
so our remaining task is to present explicit expressions 
for the transition probabilities $P_{kj}\equiv |A_{kj}|^2$.
In the adiabatic approximation of the previous section,
these $|A_{kj}|^2$ were just $\delta_{kj}$.
However, the adiabatic approximation can break down in resonance regions,
where the oscillation length increases 
by the factor $1/\sin 2\theta$ in the two-state approximation.
If the oscillation length becomes comparable to or larger than 
the scale length of the matter potential $\,|d\ln V_e(r)/dr|^{-1}$,
the medium and therefore the matter-induced mixing angles are 
changing sufficiently fast, 
so that adiabaticity is violated~\cite{Parkeformula}.
When neutrino evolution is non-adiabatic, 
$P_{kj}$ has a complicated structure and includes off-diagonal pieces.
The nature of $P_{kj}$ is discussed next.

For further progress,
we employ another piece of experimental data:
$\delta m^2_{21}$ and $\delta m^2_{32}$
differ by more than an order of magnitude,
so that both resonances in our three-neutrino system 
can be treated separately 
and involve only two levels at a time~\cite{zag}.
For each of the two resonances, 
an improved Landau--Zener approximation establishes a variant 
of the Parke formula~\cite{Parkeformula}
\beq{crossprob}
P_c=
\Theta\left(E-E_R(0)\right)\;\frac{\exp(-\Gamma\sin^2\theta)-\exp(-\Gamma)}
{1-\exp(-\Gamma)}
\eeq
as the generic leading-order form of the level-crossing probability~\cite{review}. 
Here, 
$\theta$ is the effective two-level vacuum mixing angle, 
$\Gamma$ is the ``adiabaticity parameter'', 
and $E_R(0)\equiv E_R$ is the resonance energy at the solar core. 
These quantities depend upon 
the three-level vacuum neutrino parameters, 
the matter-density profile, 
and the resonance under consideration. 
Before applying Eq.~\rf{crossprob} 
to each resonance in the present situation, 
we discuss some of its general features. 

As discussed in Sec.~\ref{subsec:diagonalizing}, 
outside the Sun's center 
the resonance energy increases $E_R(r\neq 0)>E_R$. 
It follows 
that only neutrinos produced with energies $E>E_R$ 
can experience resonance conversion
on their way to the Sun's surface. 
Since a resonance is required for appreciable level-crossing probabilities, 
we need $E>E_R$ in order for $P_c\neq 0$. 
This feature is ensured 
by the presence of the threshold function $\Theta$ in Eq.\ \rf{crossprob}. 
We also note 
that the crossing probability 
$P_c$ goes smoothly to 100\% as $\theta$ approaches zero. 
This is expected and necessary; 
it reflects the fact 
that two states cannot repel each other 
when they are decoupled. 

Explicit expressions for the adiabaticity parameter $\Gamma$ show~\cite{review} 
that it depends on the neutrino energy as $1/E$.
Thus, 
it is useful to define a ``non-adiabatic energy'' $\TEna$ 
via the simple equation 
\beq{ENAtilda}
\Gamma\equiv \frac{\TEna}{E}\,;
\eeq
$\TEna$ sets the energy scale for non-adiabatic effects. 
In the limit $E\rarr 0$, $\Gamma\rarr\infty$ and the crossing probability 
$P_c$ goes to zero.  This is the adiabatic limit.
The onset of non-adiabaticity is found by expanding 
Eq.~\rf{crossprob} under the condition $E\ll \TEna$.
One has 
\beq{lowEprob} 
P_c=\exp(-\Gamma\sin^2\theta)-\exp(-\Gamma)+\cdots\;.
\eeq 
Taking $P_c=e^{-3}\simeq 5\%$ 
as a characteristic value for the onset of non-adiabaticity, 
the leading term in Eq.\ \rf{lowEprob} determines 
an onset energy $\Ena$ for non-adiabatic effects:
\beq{onset} 
\Ena=\frac{1}{3}\sin^2\theta\,\TEna \,.
\eeq  
Thus we have an energy condition for nonadiabaticity:
$E>\max\{E_R,\,E_{NA}\}$.
Note, however, that for very large energy $E\gg\TEna$, 
$\Gamma$ goes to zero and $P_c$ approaches $\cos^2\theta$. 
This reflects the fact that at very high energies 
the oscillation wavelength exceeds the matter region, 
and so vacuum oscillations result. 

For large $\sin^2\theta$, 
the definition \rf{onset} may be inconsistent  
because the next-to-leading contribution in Eq.\ \rf{lowEprob} 
can become sizable. 
However, 
we will see below 
that only the cases $\theta=\theta_{12},\theta_{13}<45^\circ$ 
are of interest in the present context. 
The ratio of the second to first terms in the expansion \rf{lowEprob} 
evaluated at $\Ena$ 
is $\exp(-3\cot^2\theta)$. 
Even for the ultra-conservative assumption $\theta=45^\circ$, 
this ratio is less than $5\%$, 
so that Eq.\ \rf{onset} is more than sufficient for our purposes. 

\begin{figure*}[tb]
\begin{center}
\includegraphics[width=0.60\hsize]{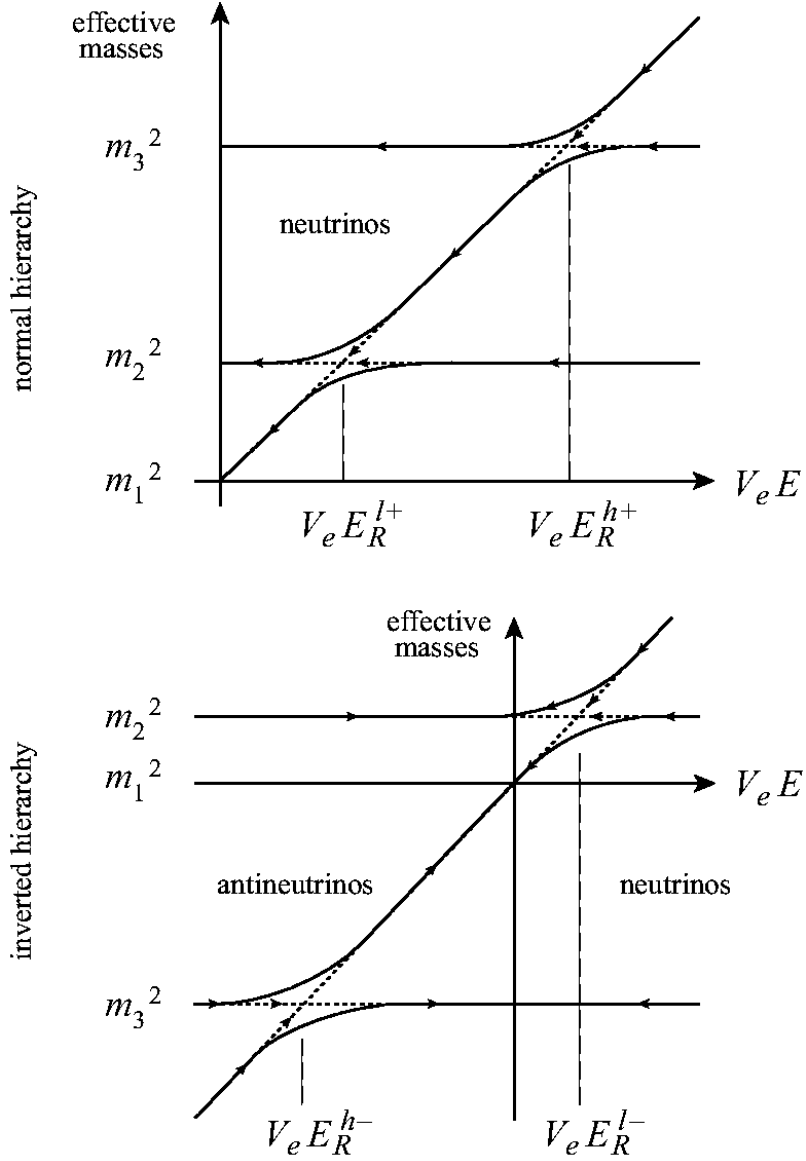}
\end{center}
\vskip-10pt
\caption{Schematic illustration of the relative positions of the lower- and
higher-energy
resonances for the normal and inverted mass hierarchies.
Neutrino (antineutrino) eigenvalues are on the right (left) of the vertical
effective-mass axis.  
The dotted lines indicate non-adiabatic transitions.}
\label{fig0} 
\end{figure*}

Although the generic crossing probability \rf{crossprob} 
provides the basis for the determination of $P_{jk}$, 
many details also depend on the number of resonances encountered (0, 1, or 2),
the mass hierarchy, 
and on whether neutrinos or antineutrinos are considered. 
For example, 
with $\theta_{13}=0$ 
the state $|3\ra$ decouples
as we have seen, 
and there is but a single resonance; 
it lies in the neutrino sector with either 
the normal or inverted mass hierarchies. 
With $\theta_{13}\ne 0$, there are two resonances;
they both lie in the neutrino sector with the normal mass hierarchy,
but lie one each in the neutrino and antineutrino sectors 
with the inverted mass hierarchy.
The schematic Fig.~\ref{fig0} displays the possibilities.
We use superscripts $l$ and $h$ on relevant variables to denote the lower- and 
higher-energy resonances, 
and $+$ and $-$ to denote the normal and inverted hierarchies, respectively. 
Some obvious energy regions where evolution is purely adiabatic are 
(i) $E_\nu \alt E^l_{NA}$, (ii) all $E_{\nubar}$ with the normal hierarchy,
(iii) all $E_{\nubar}$ with the inverted hierarchy when $\theta_{13}=0$, and 
(iv) $E_{\nubar} \alt E^h_{NA}$ with the inverted hierarchy when $\theta_{13}\neq 0$.
We will consider below the various cases, normal vs.\ inverted hierarchy,
and $\theta_{13}$ zero vs.\ nonzero. 

In principle, a neutrino produced on the back side of the Sun but traveling toward us
may encounter two resonances before leaving the production region.
There is a standard formalism available to describe this physics~\cite{review}.
However, we may ignore this ``res-in--res-out'' possibility for two reasons.
The first is that the region of DM annihilation is expected to be 
so small ($\lsim 0.01\,R_\odot$) that only 
neutrinos within a particular narrow energy band could encounter the resonances.
The second reason is that the matter density within $0.1\,R_\odot$ of the solar core
deviates from the exponential behavior in Eq.\ \rf{number_density}:
it becomes nearly constant~\cite{solar_model}. 
This near-constancy of the density profile holds even better 
within $\lambda_{\rm DM}\sim0.01\,R_\odot$.

\subsubsection{(Anti)Neutrino flavors -- normal hierarchy}
\label{subsec:normal}

\begin{figure*}[tb]
\begin{center}
\includegraphics[width=0.80\hsize]{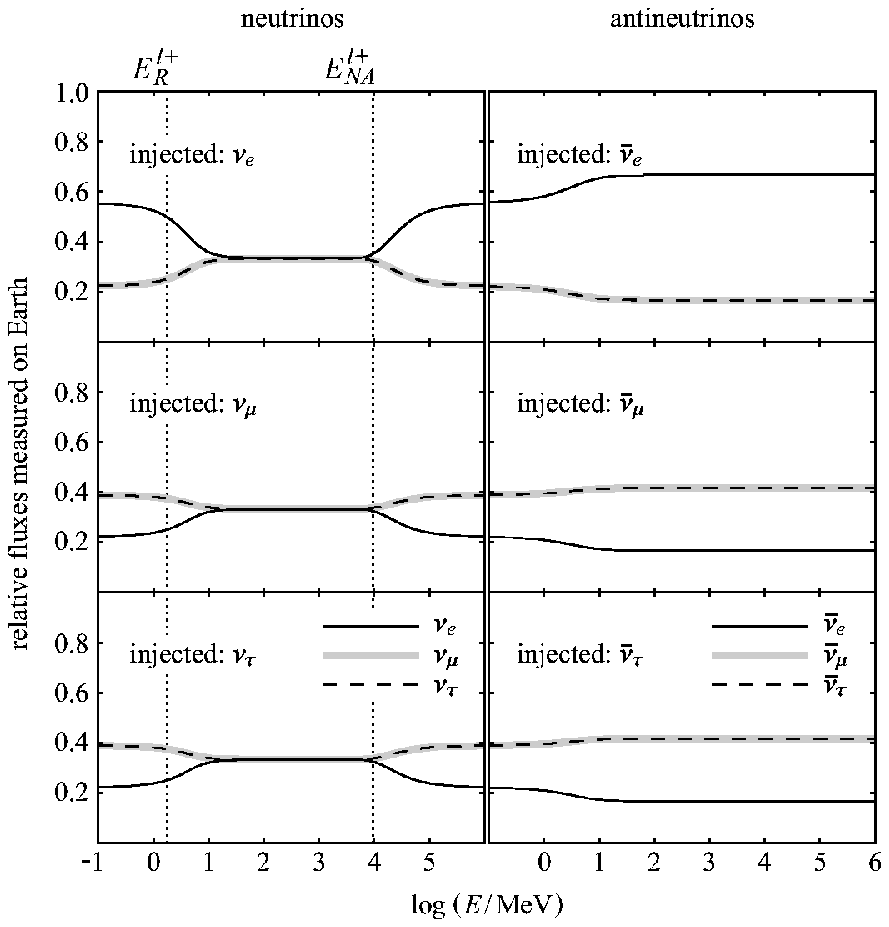}
\end{center}
\vskip-10pt
\caption{Solar neutrino and antineutrino flavor probabilities at Earth
versus energy, for a single injection flavor and for the normal mass hierarchy.
The $\nue$, $\numu$, and $\nutau$ spectra at Earth are shown as solid, gray, and 
dashed lines.
The neutrino mass-squared differences are 
$\delta m^2_{21}=8.0\times10^{-5}\,$eV${}^2$ 
and $\delta m^2_{32}=3.0\times10^{-3}\,$eV${}^2$,
and the mixing angles are
$\theta_{12}=35^{\circ}$,
$\theta_{13}=0^{\circ}$,
and $\theta_{23}=45^{\circ}$.
The CP-violating phase $\delta$ is set to zero.
The vertical dotted lines 
mark the characteristic scales 
for the lower-energy resonance 
given in Eqs.\ \rf{lowres} and \rf{nonad_L+}. 
There is no higher-energy resonance when $\theta_{13}=0$.
Note the $\numu$--$\nutau$ and $\numubar$--$\nutaubar$ interchange symmetries 
discussed in the text.
Because $\theta_{13}$ is set to zero here,
these results remain valid also for the inverted-hierarchy case.
Note that the high-energy neutrino fluxes 
are identical to the low-energy antineutrino fluxes.
This accidental feature is attributable to tribimaximal mixing, 
as explained in Appendix~\ref{accident}. 
}
\label{fig1} 
\end{figure*}

For the normal hierarchy $\delta m^2_{32}>0$,
there are two resonances for neutrinos
and none for antineutrinos.
The lower-energy neutrino resonance
occurs between the two lightest states.
For this situation, 
the characteristic quantities $E_R$, $\TEna=\Gamma\,E$, and $\theta$ 
in the crossing-probability formula~\rf{crossprob} 
are given by~\cite{review,Note1} 
\bea{lowres}
E_R^{l+} & = & 
\frac{\delta m^2_{21}\cos 2\theta_{12}}{2V_e(0)\cos^2\theta_{13}}\,,\nonumber\\
\TEna^{l+} & = &
\Gamma\,E = \pi\left|\frac{V_e}{V'_e}\right|_{r_R}
\delta m^2_{21}\,,\nonumber\\
\theta^{l+} & = & \theta_{12}\,. 
\eea
These expressions are valid at leading order. 
Appropriate to the case which we are discussing, 
the superscripts $l$ and $+$  
label the lower-energy resonance and the normal hierarchy with $\delta m^2_{32}>0$, 

The logarithmic derivative inside the absolute-value sign
is to be evaluated at the radial position  $r_R$
of the respective resonance.
In the present case of an exponential density profile,
this factor is equal to $\lambda_\odot\sim 0.1\,R_\odot$
and independent of $r$,
which simplifies the analysis.
In particular, we may set 
\beq{Ena}
\TEna^{l+} = \pi\,\lambda_\odot\,\delta m^2_{21}
     \simeq 110\,{\rm GeV}\,\frac{\delta m^2_{21}}{10^{-4}{\rm eV}^2}\,.
\eeq
The corresponding expression 
for the onset of non-adiabaticity~\rf{onset} in the present case is 
\beq{nonad_L+} 
\Ena^{l+}=\frac{1}{3}\sin^2\theta_{12}\,\TEna^{l+}\,.
\eeq 

For the higher-energy neutrino resonance
occurring between the two heaviest states,
the characteristic quantities $E_R$, $\TEna$, and $\theta$ 
are given at leading order by~\cite{review,Note2}
\bea{highres}
E_R^{h+} & = & 
\frac{\delta m^2_{32}+\delta m^2_{21}\cos^2\theta_{12}}
{2V_e(0)}\cos 2\theta_{13}\,,\nonumber\\
\TEna^{h+} & = &
\pi\left|\frac{V_e}{V'_e}\right|_{r_R}
(\delta m^2_{32}+\delta m^2_{21}\cos^2\theta_{12})\,,\nonumber\\
\theta^{h+} & = & \theta_{13}\,, 
\eea
where the superscript $h+$ 
labels the present higher-energy resonance, normal-hierarchy situation. 
From Eq.~\rf{onset}, 
we obtain here 
\beq{nonad_H+} 
\Ena^{h+}=\frac{1}{3}\sin^2\theta_{13}\,\TEna^{h+}
\eeq 
for the onset of non-adiabaticity.

\begin{figure*}[tb]
\begin{center}
\includegraphics[width=0.80\hsize]{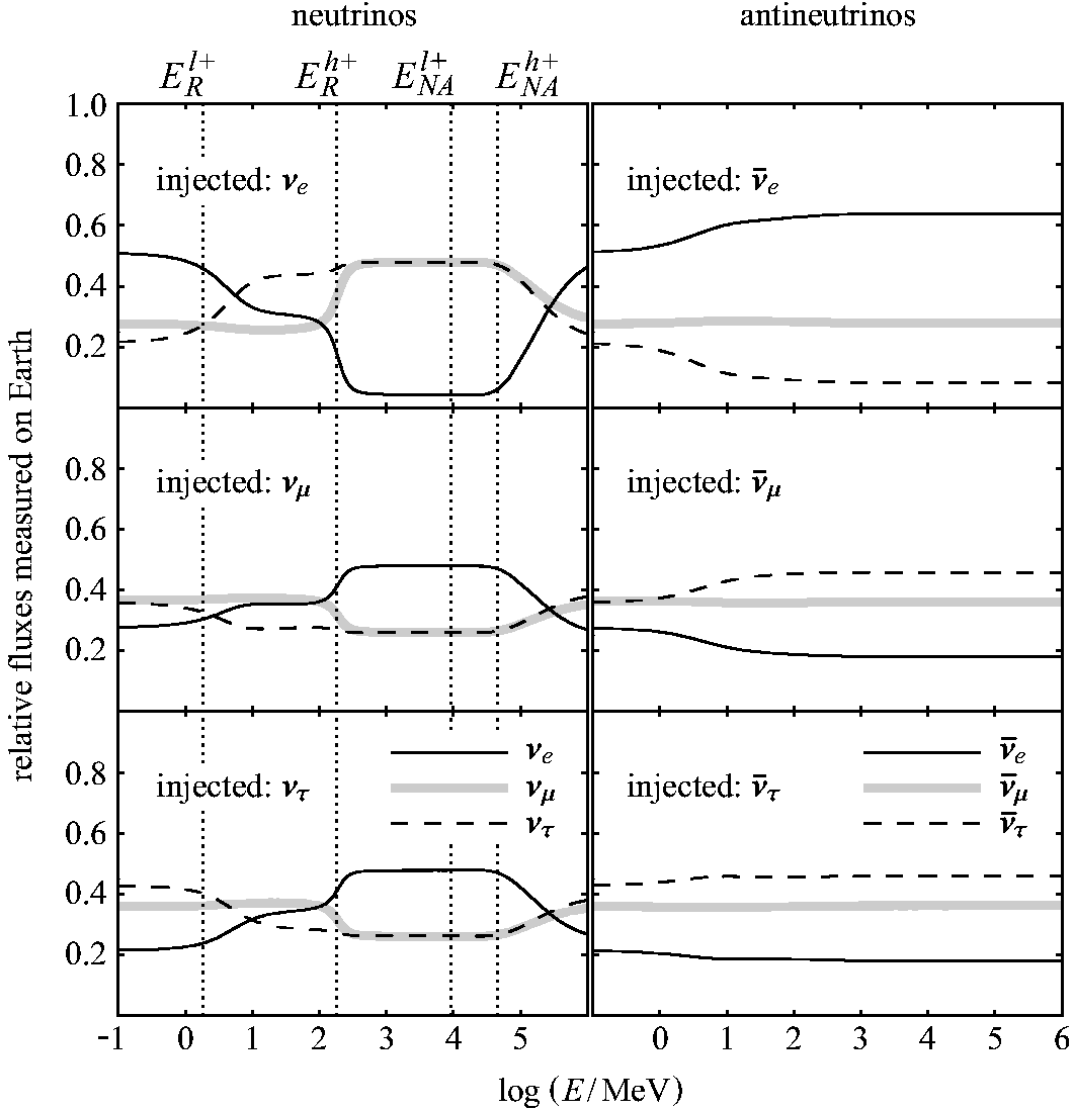}
\end{center}
\vskip-10pt
\caption{Solar neutrino and antineutrino flavor probabilities at Earth
versus energy, for a single injection flavor and for the normal mass hierarchy.
Here, we have taken $\theta_{13}=12^{\circ}$, $\delta=0$.
All other neutrino parameters
are as in Fig.\ \ref{fig1}.
The $\numu$ and $\nutau$ spectra and $\numubar$ and $\nutaubar$ spectra
are interchanged if $\delta=\pi$ is chosen.
Vertical dotted lines 
mark the characteristic scales 
for the lower-energy resonance given by Eqs.~\rf{lowres} and \rf{nonad_L+} 
and the higher-energy resonance given by Eqs.~\rf{highres} and \rf{nonad_H+}.
}
\label{fig2} 
\end{figure*} 

We now assemble all the pieces
necessary for including non-adiabatic matter effects
into our analysis for the normal-hierarchy case.
The neutrino level-crossing probabilities
are determined by
\beq{normtransprob}
P_{kj} = 
\left(\begin{array}{ccc}
1\hspace{-.5mm}-\hspace{-.6mm}P^{l}_c & P^{l}_c & 0\\
P^{l}_c & 1\hspace{-.5mm}-\hspace{-.6mm}P^{l}_c & 0\\
0 & 0 & 1
\end{array}\right)\!\!
\left(\begin{array}{ccc}
1 & 0 & 0\\
0 & 1\hspace{-.5mm}-\hspace{-.6mm}P^{h}_c & P^h_c\\
0 & P^h_c & 1\hspace{-.5mm}-\hspace{-.6mm}P^h_c
\end{array}\right)\,,
\eeq
where the superscripts $l$ and $h$
imply evaluation of the crossing probability \rf{crossprob}
at the lower resonance with Eqs.~\rf{lowres} 
and at the higher resonance with Eqs.~\rf{highres}, respectively.
We remark that with two resonances at play, 
$P_{jk}\neq P_{kj}$ in general,
so that the ordering of the two non-commuting matrices on the right-hand side 
of Eq.~\rf{normtransprob} is essential.

On the other hand, for antineutrinos there are no
resonances present in the normal-hierarchy case.
The adiabatic relation $P_{kj} = \delta_{kj}$ is exact.

The terrestrial flavor fluxes resulting from 
pure flavor injection in the Sun's center 
are plotted in Figs.~\ref{fig1} and \ref{fig2}
for representative sets of neutrino parameters. 
One sees that the processed flavors arriving at Earth are far from pure.
The neutrino spectra in Fig.~\ref{fig1} illustrates the three energy regions that 
may result from the influence of a matter resonance.
At $E_\nu\agt E^{l+}_{NA}$, the flavor evolution is complicated by non-adiabaticity.
In the region $E_R^{l+}\alt E_\nu \alt E^{l+}_{NA}$,
the flavor ratios have the simple adiabatic values presented and discussed in 
Sec.~\ref{subsec:nu-evo}.
Below $E_R^{l+}$, the matter state $|e\ra$ is not well separated from the other states,
and the description is again complicated.
A simplifying feature for all energies is the $\mu$-$\tau$~degeneracy 
that necessarily results when $\theta_{32}=45^\circ$ and $\theta_{13}=0$.

Figs.~\ref{fig1} and \ref{fig2} differ in that the minimal $\theta_{13}=0$ is input into the first, 
and the 3-$\sigma$ maximal $\theta_{13}=12^\circ$ is input into the second.
Several qualitative differences result when $\theta_{13}$ is nonzero.
The most noticeable is that with $\theta_{13}\ne 0$, 
the mixing between $|3\ra$ and an injected 
$|e\ra$ significantly suppresses the $\nue$ flavor at Earth in the adiabatic high-energy region
between $E_R^{h+}$ and $\Ena^{h+}$.
The next most noticeable change is the breaking of the $\mu$--$\tau$ degeneracy,
for both neutrinos and antineutrinos.
As explained in Sec.~\ref{subsec:delta}, the $\numu$ and $\nutau$ spectra,
and the $\numubar$ and $\nutaubar$ spectra, are interchanged if 
$\delta=0$ is replaced with $\delta=\pi$.

Although there are no antineutrino resonances with the normal hierarchy,
there are nevertheless features in the antineutrino flavor spectrum at the 
resonant energies.  These ``kinematical reflections'' are due to the following
physics:
As the energy passes from below to above the resonant value, the effective Hamiltonian~\rf{eff_ham} 
morphs from being vacuum dominated to being matter dominated.
In turn, the mixing angle(s) associated with the resonance morph from their vacuum value(s)
to their matter value(s).
If the respective signs of the vacuum and matter parts of the Hamiltonian are opposite,
as happens for neutrinos in the normal hierarchy,
then the transition may be dramatic; at resonance the mixing angle passes through $45^\circ$.
If instead, the signs of the two Hamiltonian pieces are the same,
as with antineutrinos in the normal hierarchy, then the transition in mixing-angles 
is less dramatic, but possibly still significant.
The transition of mixing angles through the lower resonant energy 
in the $\nuebar$ and $\nutaubar$ curves of Fig.~\ref{fig2} are evident.
The top antineutrino panel with pure $\nuebar$ injected is the simplest to understand.
Below resonant energies, 
the vacuum angles are relevant, 
and we expect a flavor ratio of $(5/9,\,2/9,\,2/9)$, 
as derived from Eq.~\rf{FPtribimax} 
or equivalently from Eqs.~\rf{probvacuum2a} and~\rf{probvacuum2b}.
Above the resonant energies, the injected $|\nuebar\ra$ is identified in matter 
with the state $|1\ra$. 
In vacuum (e.g., at Earth) and to lowest order in $\sin\theta_{13}\,\cos\delta$, 
this mass state decomposes into the flavor ratio 
$(2/3,\,(1+3\,\sqrt{2}\,\sin\theta_{13}\,\cos\delta)/6,
   \,(1-3\,\sqrt{2}\,\sin\theta_{13}\,\cos\delta)/6)$.
At the 3-$\sigma$ maximal value of $3\,\sqrt{2}\,\sin\theta_{13}\,\cos\delta=0.90$,
the final $\nutaubar$ fraction is driven almost to zero.

\subsubsection{(Anti)Neutrino flavors -- inverted hierarchy}
\label{subsec:inverted}
Next, we consider the inverted hierarchy,
which is characterized by $\delta m^2_{32}<0$.
Because the (decoherent) fluxes
are unaffected by the mass hierarchy when $\theta_{13}=0$,
as discussed in Sec.~\ref{subsec:diagonalizing}, Fig.~\ref{fig1} remains 
valid for either mass hierarchy.
In the more general situation with $\theta_{13}\neq0$,
it is no longer true
that the $|3\ra$ eigenstate of the Hamiltonian completely decouples,
and so a few additional considerations are necessary.
First, note that only one of the two resonances occurs for neutrinos;
the other one now appears in the antineutrino sector.
In particular,
the condition of a large resonance separation,
which affects the quality of our approximation,
is met trivially for the inverted hierarchy.
Second,
for the analysis of each resonance
we can employ the same bases
and approximations
as we used for the normal hierarchy
because the analysis essentially depends on the experimental result
that $|\delta m^2_{32}|\gg|\delta m^2_{21}|$ 
independent of the sign of $\delta m^2_{32}$.

Armed with these considerations, we find 
that the resonance between the states $|1,r\rangle$ and $|2,r\rangle$
stays in the neutrino sector, 
and the present results are equal to those in the lower-resonance normal-hierarchy situation. 
Hence, we only need to replace the superscripts 
$l+\to l-$
in Eqs.\ \rf{lowres} and \rf{nonad_L+}. 
This result is expected 
since the analysis of the 1--2 resonance 
involves neither $|3,r\rangle$
nor the value of $\delta m^2_{32}$.
The resulting jump probability for neutrinos is 
\beq{invtransprob}
P_{kj} = 
\left(\begin{array}{ccc}
1\hspace{-.5mm}-\hspace{-.6mm}P^l_c & P^l_c & 0\\
P^l_c & 1\hspace{-.5mm}-\hspace{-.6mm}P^l_c & 0\\
0 & 0 & 1
\end{array}\right),
\eeq
where $P_c$ is evaluated with Eqs.\ \rf{crossprob}, \rf{lowres}, and 
the replacement $l+\to l-$.
Note that $|1,r\rangle$ and $|2,r\rangle$
are now the two heaviest states.

\begin{figure*}[tb]
\begin{center}
\includegraphics[width=0.80\hsize]{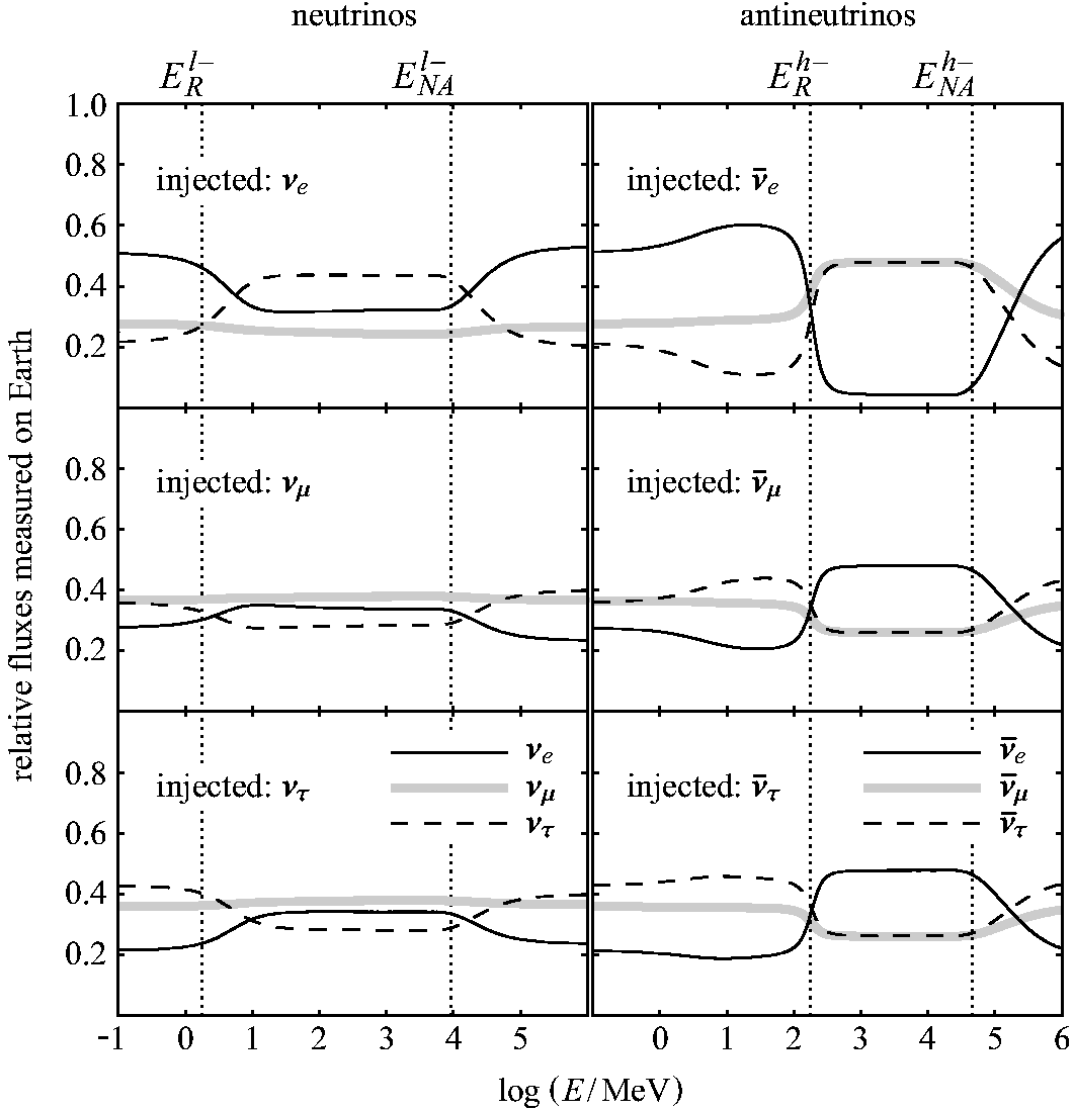}
\end{center}
\vskip-10pt
\caption{Neutrino and antineutrino flavor probabilities on Earth
versus energy, for the inverted hierarchy.
Here, we have taken $\delta m^2_{32}=-3.0\times10^{-3}\,$eV${}^2$.
All other neutrino parameters
are as in Fig.~\ref{fig2} (including $\theta_{13}=12^\circ$ and $\delta=0$).
The $\numu$ and $\nutau$ spectra and $\numubar$ and $\nutaubar$ spectra 
are interchanged if $\delta=\pi$ is chosen.
}
\label{fig3} 
\end{figure*}

For antineutrinos,
the resonance involves $|1,r\rangle$ and $|3,r\rangle$,
which are the two lightest states in the inverted mass hierarchy.
An argument paralleling that leading to Eq.~\rf{highres} yields
for this case
\bea{antires}
E_R^{h-} & = & 
-\frac{\delta m^2_{32}+\delta m^2_{21}\cos^2\theta_{12}}
{2V_e(0)}\cos 2\theta_{13}\,,\nonumber\\
\TEna^{h-} & = &
\Gamma\,E = \pi\left|\frac{V_e}{V'_e}\right|_{r_R}
(-\delta m^2_{32}-\delta m^2_{21}\cos^2\theta_{12})\,,\nonumber\\
\theta^{h-} & = & \theta_{13}\,.
\eea
Here, 
the superscript $h-$ denotes the higher-energy inverted-hierarchy resonance. 
The energy for the onset of non-adiabaticity is given by 
\beq{nonad_H-} 
\Ena^{h-}=\frac{1}{3}\sin^2\theta_{13}\,\TEna^{h-}\, ,
\eeq 
and the jump-probability matrix now takes the form
\beq{invantitransprob}
P_{kj} = 
\left(\begin{array}{ccc}
1\hspace{-.5mm}-\hspace{-.6mm}P^h_c & 0 & P^h_c\\
0 & 1 & 0\\
P^h_c & 0 & 1\hspace{-.5mm}-\hspace{-.6mm}P^h_c
\end{array}\right).
\eeq
Here,
 $P^h_c$ is determined by Eqs.\ \rf{crossprob} and \rf{antires}.
The inverted-hierarchy flavor fluxes
on Earth
for a representative set of neutrino parameters 
are shown in Fig.~\ref{fig3}.
A comparison of Figs.~\ref{fig2} and \ref{fig3} reveals that 
the effect of moving the higher-energy resonance from the neutrino to the 
antineutrino sector via the change in mass hierarchy is dramatic.

The density-matrix formalism presented above
accommodates any mixture of flavor fluxes produced in the solar core.
For illustration, 
we have selected only examples 
with pure single-flavor initial states 
in Figs.~\ref{fig1}, \ref{fig2}, and \ref{fig3}. 
The previously discussed sources are effectively incoherent, 
which implies
that the fluxes for mixed-flavor cases
can be inferred from these figures
by simply taking the appropriately weighted incoherent sum
of our pure-flavor results.

\subsubsection{Sensitivity to ${\theta_{13}}$} 

Because $\theta_{13}$ is experimentally bounded 
to be less than about $12^\circ$, 
its measurement will be more difficult 
than that of the other two mixing angles. 
It is therefore interesting 
to discuss the $\theta_{13}$ sensitivity of the terrestrial flavor ratios 
of neutrinos from solar WIMP annihilation. 

Comparison of Fig.~\ref{fig1} 
with Figs.~\ref{fig2} and \ref{fig3} reveals 
that the relative flavor fluxes on Earth 
indeed depend on $\theta_{13}$. 
The normal-hierarchy antineutrinos in Fig.\ \ref{fig2} 
and the inverted-hierarchy neutrinos in Fig.\ \ref{fig3} 
show changes with respect to Fig.\ \ref{fig1} 
that are of the expected size $\sin 12^\circ\sim 0.2$. 
However, 
the changes for normal-hierarchy neutrinos 
and inverted-hierarchy antineutrinos 
are more significant. 
The question arises 
as to why a relatively small change in $\theta_{13}$ 
can lead to effects of this magnitude. 

The answer to this question 
lies in the transition from adiabaticity to non-adiabaticity 
for the higher resonance, 
which is governed by $\theta_{13}$: 
Equations~\rf{nonad_H+} or~\rf{nonad_H-} show 
that the onset of non-adiabaticity directly scales with $\sin^2\theta_{13}$. 
In Fig.\ \ref{fig1}, 
we have taken $\theta_{13}=0$. 
The loss of adiabaticity therefore occurs at zero energy. 
Moreover, the level-crossing probability~\rf{crossprob} becomes $P^h_c=1$ for zero mixing angle. 
In other words, 
the $|3\rangle_m$ state is decoupled 
and does  not participate in the adiabatic label permutation 
resulting from level repulsion, 
as discussed in Sec.\ \ref{subsec:diagonalizing}.
In Figs.\ \ref{fig2} and \ref{fig3}, 
however, 
$\theta_{13}\neq 0$
and $|3\rangle_m$ is not decoupled. 
If in such a situation 
an adiabatic region exists---as in the case for $\sin\theta_{13}=12^\circ$---
the state experiences level permutation $|3\rangle_m\to|2\rangle_m$ 
while propagating to the solar surface. 
It is this mechanism 
that causes the large $\theta_{13}$ effects 
for normal-hierarchy neutrinos 
and inverted-hierarchy antineutrinos. 

How small can $\theta_{13}$ be 
for this effect to occur? 
Since the label permutation 
is valid only for energies above resonance 
in the adiabatic regime, 
we must require 
that $E$ lies above the higher resonance  
and below the onset of non-adiabaticity for this resonance, 
i.e.,  
\bea{conditions} 
E^{h+}_R < &E& <\Ena^{h+}\, ,\nonumber\\
E^{h-}_R < &E& <\Ena^{h-}\, ,
\eea
for 
normal-hierarchy neutrinos 
and inverted-hierarchy antineutrinos, 
respectively. 
Here, 
the various energy scales 
are given by Eqs.\ \rf{highres}, \rf{nonad_H+}, 
\rf{antires}, and \rf{nonad_H-}. 
It follows 
that the onset of non-adiabaticity must 
lie above the resonance energy. 
Such will be the case 
for values of $\theta_{13}$ down to a fraction of a degree. 

We conclude 
that the terrestrial flavor fluxes from neutrinos originating in 
solar WIMP decay  
are quite sensitive to the mixing angle $\theta_{13}$. 
Accordingly, the study of solar-WIMP annihilation properties 
would greatly benefit from a pre-knowledge of the size of $\theta_{13}$. 
On the other hand, 
if dark-matter properties 
can be extracted from other experiments, 
such as direct WIMP discovery in underground detectors,
or even better, at the LHC or ILC, then 
a measurement of $\theta_{13}$ 
through WIMP annihilation in the Sun 
may be feasible.

\subsubsection{The CP-violating phase ${\delta}$} 
\label{subsec:delta}
Currently, 
the CP-violating phase $\delta$ is unconstrained by experiments.  
We are thus led to study the effects of $\delta$ 
on the propagation of high-energy solar neutrinos. 

The general non-adiabatic density-matrix framework 
discussed in  
Sec.\ \ref{subsec:nonadiabatic} 
has made no assumptions 
regarding the size of the phase $\delta$. 
The energy range of interest for solar neutrinos from WIMP decay 
is a few $100\,$MeV to about $100\,$GeV. 
Below this range, 
matter effects are suppressed 
and the standard vacuum results are approached. 
Above this range, 
absorption effects become dominant, 
and our framework is no longer accurate. 
Since this energy range is mainly below $\Ena^{h\pm}$
and $\Ena^{l\pm}$, 
the adiabatic approximation is useful. 

We again take $\theta_{12}$ and $\theta_{32}$ 
close to their observed tribimaximal values 
given in Sec.~\ref{subsec:vacuumoscillations}. 
For the remaining mixing angle, 
the largest experimentally allowed value $\theta_{13}=12^\circ$ 
is most interesting:
From Eq.~\rf{R13complex} one may infer that in a CP-conserving calculation,
as here for decohered neutrino propagation,
the $\delta$ dependence will always occur in the combination 
$\Re{(e^{i\delta}\sin\theta_{13})}=\cos\delta\,\sin\theta_{13}$.
It is therefore only in combination with larger values of $\theta_{13}$ that one may seek 
observable $\delta$ effects.
Thus, when illustrating effects of nonzero $\delta$, we will generally 
dramatize the result by taking $\theta_{13}$ to be $12^\circ$, 
its maximally allowed 3-$\sigma$ value.

Concerning the $\delta$-parameter,
two important analytic features emerge.
The first is {\bf Feature~(i)}: 
{\it 
For $\theta_{32}=45^\circ$, but arbitrary $\theta_{21}$ and $\theta_{13}$,
the replacement $\delta\rightarrow \pi-\delta$ 
interchanges the role of $\nu_{\mu}$ and $\nu_{\tau}$,
and $\numubar$ and $\nutaubar$,
and leaves unaffected the terrestrial $\nue$ and $\nuebar$ fluxes.} 
Although some versions of this result are known, 
we have included a proof in Appendix~\ref{proof1} for completeness. 
We remark that Feature (i) makes no assumptions about the neutrino energy
or adiabaticity,
so that its validity actually extends beyond our energy range of interest,
from a few $100\,$MeV to about $100\,$GeV. 

Assuming continuity, 
Feature (i) implies that for $\delta=90^\circ$, 
the $\nu_{\mu}$ and $\nu_{\tau}$ fluxes are identical. 
This is consistent 
with the previously determined condition $\Re (U_{e3})=0$
for $\nu_{\mu}$--$\nu_{\tau}$ interchange symmetry: 
inspection of the mixing matrix $U$ shows 
that $U_{e3}=-i\sin\theta_{13}$ is purely imaginary for $\delta=90^\circ$. 

The second result is {\bf Feature~(ii)}: 
{\it For $\theta_{32}$ and $\theta_{12}$ at their tribimaximal values, 
the terrestrial flavor ratios for normal-hierarchy neutrinos 
and inverted-hierarchy antineutrinos 
are independent of $\delta$ to zeroth order in 
$(\delta m^2_{21}/\delta m^2_{32})$.
In addition, the terrestrial $\nu_e$ and $\overline{\nu}_e$ fluxes 
are independent of $\delta$ to zeroth order in 
$(\delta m^2_{21}/\delta m^2_{32})$ for any hierarchy.} 
This feature holds only in a finite energy region 
determined by adiabaticity 
and by $V_eE\gg\delta m^2_{jk}$. 
Fortunately, this energy region coincides 
with our range of interest, from a few $100\,$MeV to about $100\,$GeV;
and fortunately again, $(\delta m^2_{21}/\delta m^2_{32})\sim 1/30$ is a small parameter.
We verify this result in Appendix~\ref{proof2}. 

As mentioned above, 
$\theta_{13}=0$ eliminates $\delta$ from the mixing matrices $U$ and $U_m$. 
It follows 
that Fig.\ \ref{fig1}, 
for instance, 
is unaffected by the value of $\delta$.  
Figures \ref{fig2} and \ref{fig3}, 
on the other hand, 
should exhibit some changes 
when $\delta$ is varied. 
With Features~(i) and (ii) at hand, 
certain phase effects on the terrestrial flavor fluxes 
can be understood qualitatively. 
An immediate consequence of Features~(i) and (ii) is 
the degeneracy of the $\nu_{\mu}$ and $\nu_{\tau}$ flavor spectra 
in the adiabatic region 
for normal-hierarchy neutrinos 
and inverted-hierarchy antineutrinos. 
This characteristic is apparent in Figs.~\ref{fig2} and \ref{fig3}. 
Feature (ii) implies that $\delta$ effects 
are confined to the $\overline{\nu}_\mu$ and $\overline{\nu}_\tau$ fluxes 
in Fig.~\ref{fig2} and to the $\nu_\mu$ and $\nu_\tau$ fluxes 
in Fig.~\ref{fig3}. 
We must look at these fluxes when seeking 
any sizable $\delta$ dependence. 

The qualitative behavior of these fluxes 
under changes of $\delta$ 
is dictated by Feature (i). 
As $\delta\to90^\circ$,  $\delta$ and $\pi-\delta$ approach each other, 
the $\mu$- and $\tau$-flavor fluxes must move toward each other; 
at $\delta=90^\circ$ they become degenerate. 
When $\delta$ is increased further, 
the $\mu$- and $\tau$-flavor fluxes 
reverse their role 
with respect to each other. 
The first-order results for $A_0$ 
in Appendix~\ref{proof2} establish 
that the $\delta$ dependence of these fluxes 
is of the generic form $B+C\sin\theta_{13}\cos\delta$, 
where $B$ and $C$ depend on the mass hierarchy, 
on $\nu$ versus $\nubar$, and on $\theta_{13}$. 
The precise expressions for $B$ and $C$ 
in each case can be inferred from Eqs.\ \rf{R0_anti_normal} and \rf{R0_neu_inv}. 

\begin{figure*}[tb]
\begin{center}
\includegraphics[width=0.80\hsize]{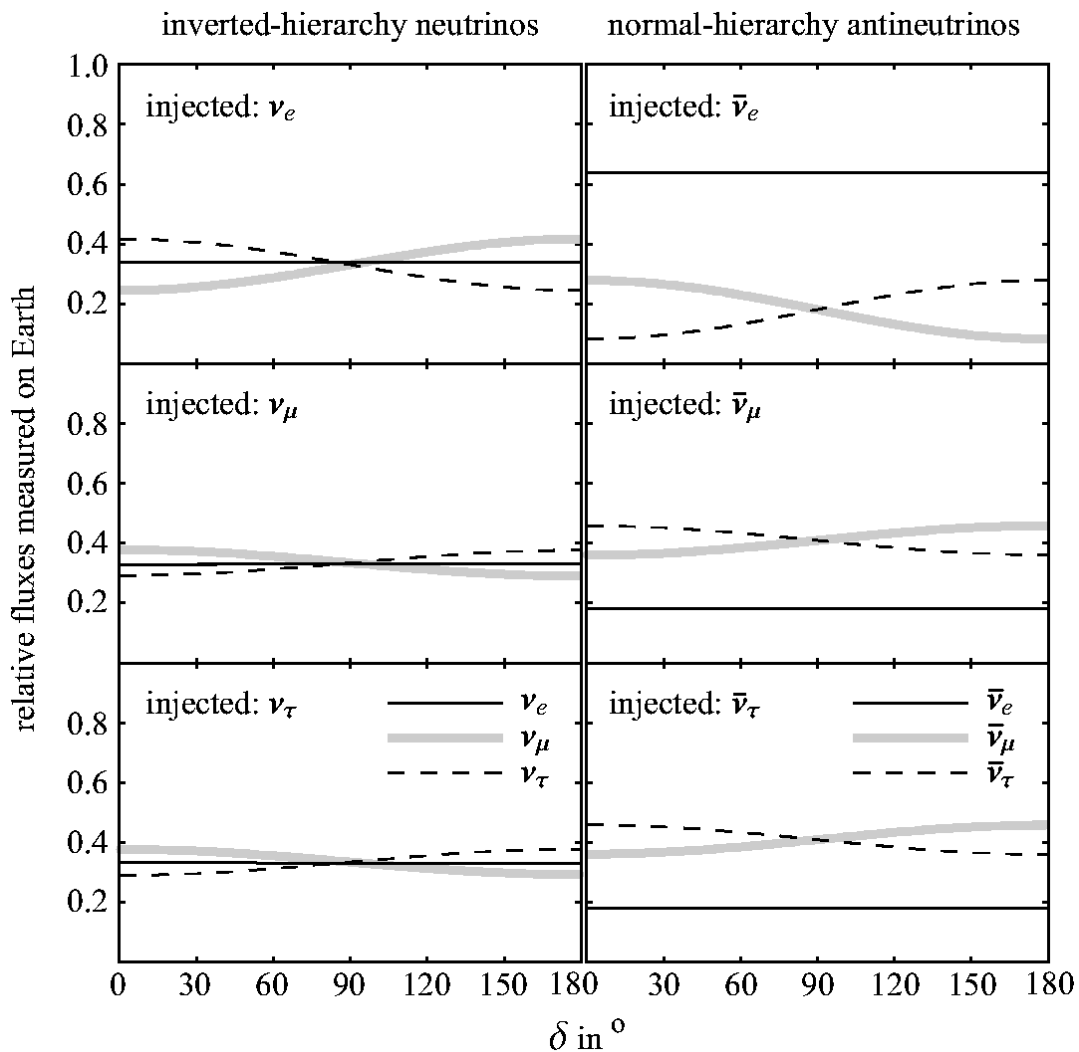}
\end{center}
\vskip-10pt
\caption{
Terrestrial neutrino fluxes versus the phase $\delta$. 
The neutrino mass-squared differences are 
$\delta m^2_{21}=8.0\times10^{-5}\,$eV${}^2$ 
and $\delta m^2_{32}=\pm3.0\times10^{-3}\,$eV${}^2$,
where the upper (lower) sign 
refers to the normal (inverted) hierarchy. 
The mixing angles are
$\theta_{12}=35^{\circ}$,
$\theta_{13}=12^{\circ}$,
and $\theta_{23}=45^{\circ}$, 
and the neutrino energy is $E=10\,$GeV. 
The $\mu$- and $\tau$-flavor fluxes 
exhibit opposite $\cos\delta$ dependences, 
whereas the $\nue$-flavor fluxes are approximately constant.  
Curves for normal-hierarchy neutrinos and inverted-hierarchy antineutrinos
are not shown, as they are all nearly constant as a function of $\delta$. 
These various results illustrate Features (i) and (ii) derived in the text.
}
\label{fig4} 
\end{figure*}

In Fig.~\ref{fig4}, 
the $\delta$ dependences of the various fluxes for inverted hierarchy (IH) neutrinos and 
normal hierarchy (NH) antineutrinos are shown.
The NH neutrino and IH antineutrino spectra are not shown  
since their $\cos\delta$ dependence is suppressed by the 
factor $(\delta m^2_{21}/\delta m^2_{32})\sim 1/30$.
Although the neutrino energy in the figure is taken to be $E=10\,$GeV, 
the results remain approximately unchanged as 
long as $E$ stays within our selected energy range. 
Apparent in Fig.~\ref{fig4} are the features discussed above. 
In particular, the $\cos\delta\lrarr -\cos\delta$ equivalence 
to the $\mu\lrarr\tau$ flavor interchange is evident.

\section{WIMP-model predictions for earthly neutrino flavors}
\label{sec:models}
In general, the WIMP LSP is a superposition of four Majorana neutralino states.
These are the two neutral Higgsinos (SUSY partners of the two Higgs doublets), the Bino
(SUSY partner of the $SU(2)$-singlet $B_\mu$ vector boson), and the neutral 
Wino (SUSY partner of the neutral $W_\mu^0$ member of the $SU(2)$-triplet gauge boson).
WMAP data yield $\Omega_{\rm DM}\simeq 23\%$, 
which in turn constrains 
the decoupling of the WIMP 
and therefore bounds the WIMP mass and superposition.
The non-observation of WIMPs in Tevatron data at Fermilab further constrains combinations of 
WIMP mass and superposition.
However, the fact remains that almost any neutralino superposition 
can be made consistent with present limits~\cite{Xman}.
Theoretical arguments are largely expressions of preference or prejudice.
For example, proponents of ``gauge-mediated SUSY breaking'' are led to 
a ``Bino-like'' LSP, while proponents of ``anomaly-mediated SUSY breaking''
are led instead to a ``Wino-like'' LSP.
Thus, there is little guidance as to what neutralino superposition comprises the LSP.

Fortunately, two disjoint sets of annihilation channels are common in LSP models.  
One (I) is the $W^+\,W^-$ channel.
The other (II) is the $b\,\bbar\, +\,\tau^+\,\tau^-$ channel.
There are further channels available to WIMP annihilation beyond (I) and (II),
but these are calculated to be subdominant.
Higgsino and Wino annihilation favor channel I, if the LSP mass exceeds $M_W$;
Higgsino annihilation but not Wino annihilation also produces $Z\,Z$ if above threshold,
but the neutrino spectrum resulting from this channel is similar to that from $W^+\,W^-$,
so we need not consider $Z\,Z$ separately.
Bino annihilation favors channel II, although even a small amount of mixing of the 
Bino with other neutralinos is likely to lead to a dominance of $W^+\,W^-$ production. 
For an LSP with mass less than $M_W$, the decay channel of the LSP is (II).

Since the annihilating WIMPS are a CP-symmetric initial state,
the neutrino and antineutrino spectra emerging from WIMP annihilation 
must be identical. The flavor processing by solar matter, described in this work,
is not CP-symmetric.  Therefore, differences between neutrino and antineutrino 
flavor ratios at Earth are expected.
Conversely, any observed differences between neutrino and antineutrino 
flavor ratios at Earth are an indication of the solar matter effect.
In principle, it is possible to separate neutrino and antineutrino spectra 
in proposed, large, magnetized iron-calorimeter experiments.

It is not hard to understand the reasons for the two preferred decay channels.
Pairs of $SU(2)$-doublet Higgsinos or  Winos have an electroweak coupling to $W^+\,W^-$ 
suppressed only by the $t$-channel $\tilde{W}$ exchange.
However, the $SU(2)$-singlet Bino does not have this coupling.
On the other hand, when the $W^+\,W^-$ channel is not available, all LSP pairs, being Majorana
particles, annihilate to massless final-state pairs in an $l=1$ ($p$-wave) angular-momentum state 
suppressed by $\beta_{\rm DM}^3$, where $\beta_{\rm DM}$ is the velocity of the LSP\@.
With massive final-state pairs, chirality flips enable an $s$-wave decay suppressed by the 
factor $\beta_{\rm DM}\,(m^2_{\rm final}/M^2_{\rm DM})$.  
The velocity of the non-relativistic LSPs 
is sufficiently small that the $s$-wave dominates and one gets the 
$b\,\bbar$ final state $N_C\,m^2_b/(N_C\,m^2_b+m^2_\tau)=95\%$ of the time, 
and the $\tau^+\tau^-$ final state $m^2_\tau/(N_C\,m^2_b+m^2_\tau)=5\%$ of the time;
here, $N_C=3$ is the number of QCD colors.
The neutrinos from tau decay $\tau\rarr \nutau+\cdots$ are more copious and 
provide a harder spectrum than those from $b$ decay, so it turns out that 
the 5\% $\tau^+\,\tau^-$ mode makes a significant contribution to the final neutrino
flux.

\begin{figure*}[tb]
\begin{center}
\includegraphics[width= 0.80\hsize]{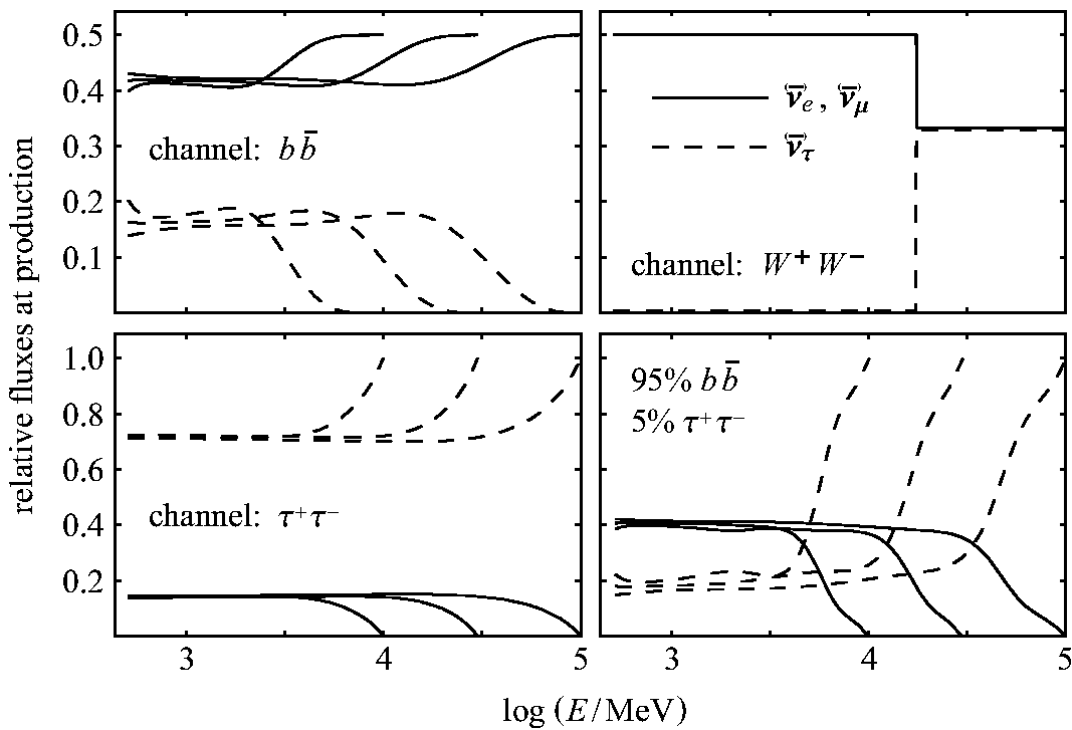}
\end{center}
\vskip-10pt
\caption{
Relative (anti)neutrino flavor fluxes at production in the solar core
are plotted  versus neutrino energy $E$ for four scenarios: 
dark-matter annihilation predominantly into $b\overline{b}$, 
or predominantly into $W^+W^-$, 
or predominantly into $\tau^+\tau^-$, 
or predominantly into 95\% $b\,\overline{b}$ + 5\% $\tau^+\tau^-$. 
Solid (dashed) lines display the $e$ and $\mu$ ($\tau$) flavor fluxes.
For each scenario, dark-matter masses $m_{\rm DM}$ of 
$10\,$GeV, $30\,$GeV, and $100\,$GeV have been considered. 
The dark-matter mass 
can be inferred from the threshold energies of the graphs.
This plot was generated from the results in Ref.~\cite{Cirelli05}.}
\label{fig5} 
\end{figure*}

In the figures to follow, 
in which all $w_{\alpha}$ are taken from Ref.~\cite{Cirelli05}, 
we will display the flavor ratios to be expected at Earth 
for the two WIMP annihilation classes (I) and (II).
For further insight, we will also show separately results for the pure 
$b\,\bbar$ and pure $\tau^+\,\tau^-$ decay modes.
However, before displaying the flavor ratios expected at Earth, we show in Fig.~\ref{fig5}
the flavor ratios produced by WIMP annihilation in the Sun for each of the four 
decay modes.
These flavor spectra are qualitatively easy to understand.
The $b\,\bbar$ production and subsequent decay to (charm+) $\tau$+$\nutau$ 
is phase-space suppressed relative to decay 
to  (charm+) e+$\nue$ or $\mu$+$\numu$,
particularly so when the neutrino is produced at higher energy.
In $\tau^+\,\tau^-$ production and decay, the branching ratio to $\nutau$ is unity,
but there are also 17\% branching ratios to $\nue$ and to $\numu$.
Accompanying the $\numu$ is a muon, which will itself decay to produce another 
$\numu$ and $\nue$, but at much lower energy since the muon loses considerable energy in the Sun's magnetic field
before decaying.
So roughly, the flavor ratios from the $\tau^+\,\tau^-$ mode are $(1.0 : 0.17 : 0.17)/1.34$.
In $W^+\,W^-$ production and decays, neutrinos at high energies are democratically produced,
and the associated taus and muons quickly lose energy in the Sun's center.
However, at lower energy, the neutrino flux originates from the decay of nearly-stopped heavy quarks
produced in $W$ fragmentation, so there is a low-energy depletion of $\nutau$s.

A useful summary of Fig.~\ref{fig5} is that, at production in the Sun, 
(i) the flavor fluxes are the same for neutrino and antineutrino,
and
(ii) the $e$ and $\mu$ flavors are produced in equal amounts, but the 
amount of $\tau$ flavor depends on the WIMP annihilation channel.
In the  $b\,\bbar$ channel, $\tau$ is suppressed relative to $e$ and $\mu$,
especially at higher energies.
In the $\tau^+\,\tau^-$ channel, it is just the opposite, with 
the $\tau$-flavor contribution enhanced, especially at higher energies.
The implication for the realistic channel (II) consisting of  
95\%~$b\,\bbar\,+\,$5\%~$\tau^+\,
\tau^-$ is a suppressed $\tau$ component at lower energies, and a highly enhanced $\tau$ 
component at higher energies.
For the realistic $W^+\,W^-$ channel (I), 
the $\tau$ component is negligible below about 20~GeV,
whereas $\tau$, $e$, and $\mu$ are equally represented above 20~GeV.
It was claimed in Sec.~\ref{subsec:nonadiabatic} that a democratic $1:1:1$
flavor ratio at production necessarily implies a democratic flavor ratio 
at Earth.  Thus, we expect flavor democracy at Earth for neutrino energies above $\sim 20$~GeV,
if the WIMP annihilation mode in the Sun is dominantly $W^+\,W^-$.
Figures to be shown will bear this out.
For all other channels, and for the $W^+\,W^-$ channel below 20~GeV,
the flavors at production are far from democratic,
and the flavor-evolution machinery developed in this work 
is important.

\begin{figure*}[tb]
\begin{center}
\includegraphics[width=0.95\hsize]{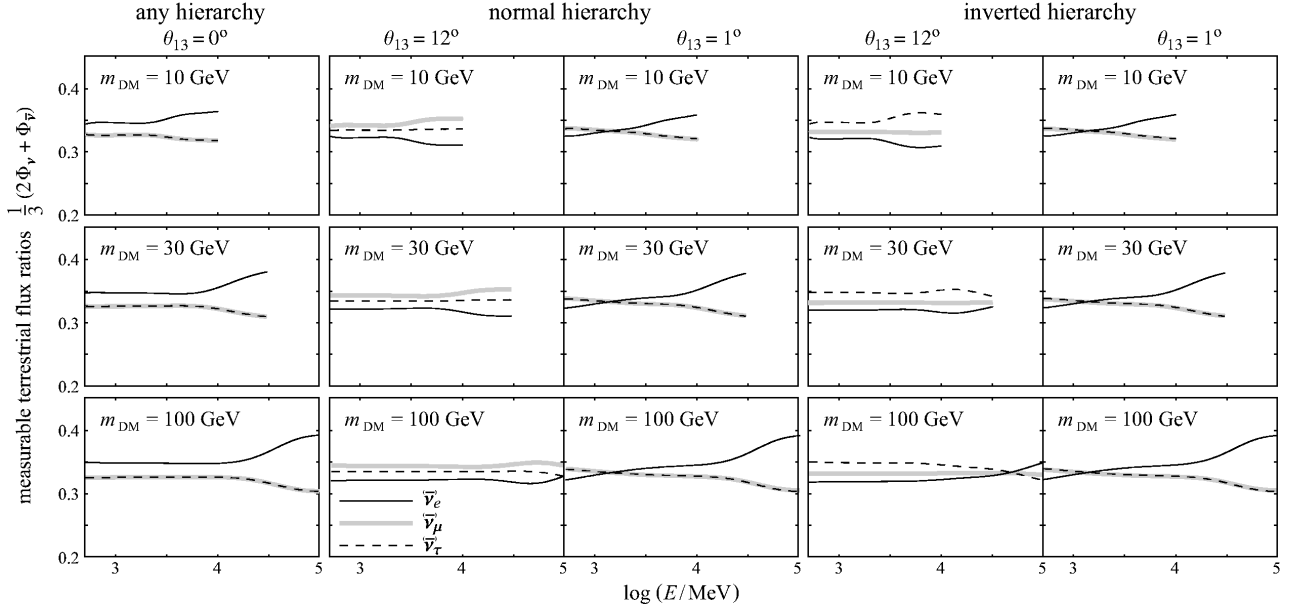}
\end{center}
\vskip-10pt
\caption{
Terrestrial fluxes versus neutrino energy $E$ 
for solar WIMP annihilation predominantly into $b\,\overline{b}$. 
The quantity $(2\Phi_\nu+\Phi_{\overline{\nu}})/3$ is plotted, 
where the $\nu$ and $\nubar$ fluxes are weighted according to their cross sections for detection.
This quantity would be measured by Earth-based detectors 
that cannot resolve $\nu_\alpha$ from $\overline{\nu}_\alpha$ events, 
as discussed in Sec.\ \ref{together}. 
The neutrino mass-squared differences are 
$\delta m^2_{21}=8.0\times10^{-5}\,$eV${}^2$ 
and $\delta m^2_{32}=\pm 3.0\times10^{-3}\,$eV${}^2$.
The mixing angles are
$\theta_{12}=35^{\circ}$ and $\theta_{23}=45^{\circ}$, 
and the CP-violating phase $\delta$ has been set to zero. 
We have considered the values $\theta_{13}=0^{\circ}$, $1^{\circ}$, and $12^{\circ}$ 
for each hierarchy 
and for each of the dark-matter masses $m_{\rm DM}=10\,$GeV, $30\,$GeV, and $100\,$GeV 
(from top to bottom).
The $\numu$ and $\nutau$ spectra as well as the $\numubar$ and $\nutaubar$ spectra
are interchanged if $\delta=\pi$ is chosen.
Note that the observables depend noticeably on the mixing angle $\theta_{13}$. 
Moreover, for larger $\theta_{13}$ angles 
there are visible differences between the normal and inverted hierarchies. 
}
\label{fig6} 
\end{figure*}

\begin{figure*}[tb]
\begin{center}
\includegraphics[width=0.95\hsize]{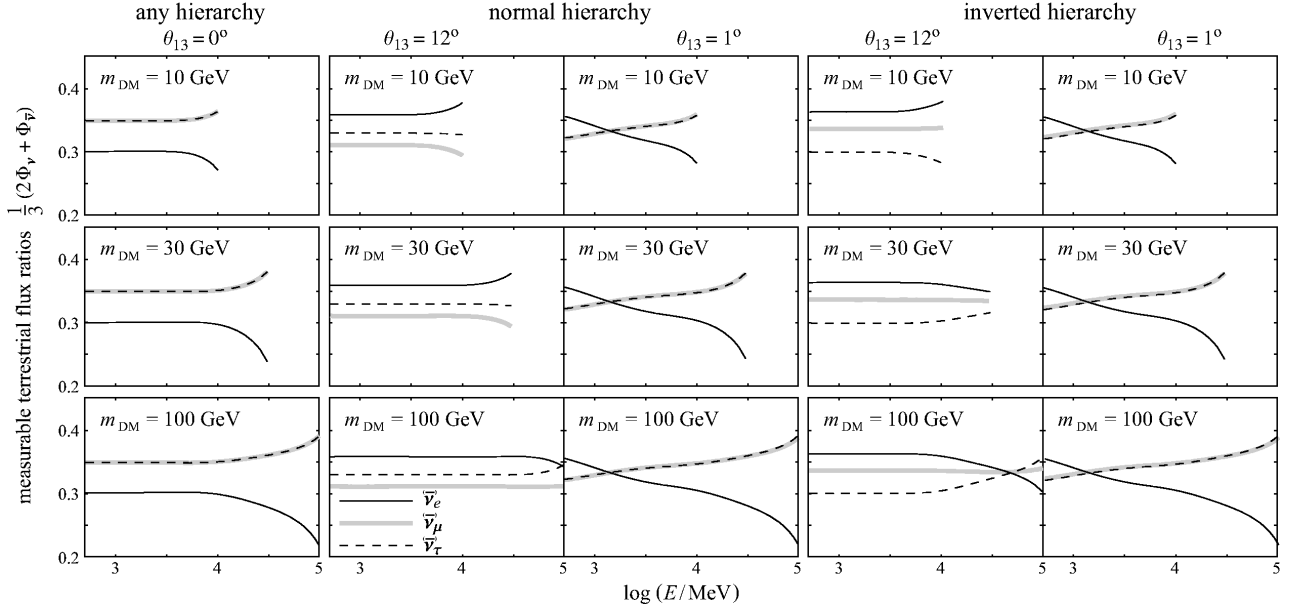}
\end{center}
\vskip-10pt
\caption{
Terrestrial fluxes versus neutrino energy $E$ 
for solar WIMP annihilation predominantly into $\tau^+\tau^-$. 
Again, the observable $(2\Phi_\nu+\Phi_{\overline{\nu}})/3$ is plotted. 
Neutrino parameters are identical to those 
employed in Fig.\ \ref{fig6}. 
The $\numu$ and $\nutau$ spectra as well as the $\numubar$ and $\nutaubar$ spectra
are interchanged if $\delta=\pi$ is chosen. 
Sensitivities to $\theta_{13}$ and the mass hierarchy 
occur at levels similar to those 
of the $b\overline{b}$ case in Fig.~\ref{fig6}. 
}
\label{fig7} 
\end{figure*}

\begin{figure*}[tb]
\begin{center}
\includegraphics[width=0.95\hsize]{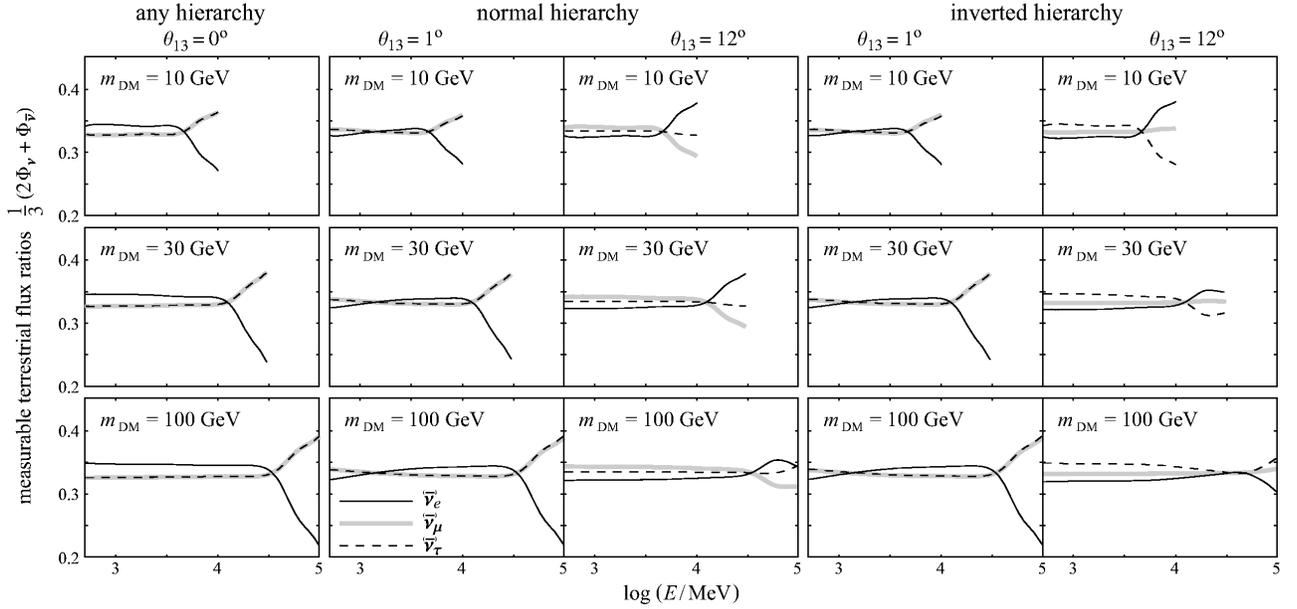}
\end{center}
\vskip-10pt
\caption{
Terrestrial fluxes versus neutrino energy $E$ 
for solar WIMP annihilation into 95\% $b\overline{b}$ + 5\% $\tau^+\tau^-$. 
Again the observable $(2\Phi_\nu+\Phi_{\overline{\nu}})/3$ is plotted, 
and neutrino parameters are identical to those in Figs.~\ref{fig6} and \ref{fig7}. 
The $\numu$ and $\nutau$ spectra as well as the $\numubar$ and $\nutaubar$ spectra
are interchanged if $\delta=\pi$ is chosen. 
Sensitivities to $\theta_{13}$ and the mass hierarchy 
are comparable to those in Figs.~\ref{fig6} and \ref{fig7}, 
as expected. 
}
\label{fig8} 
\end{figure*}

\begin{figure*}[tb]
\begin{center}
\includegraphics[width=0.95\hsize]{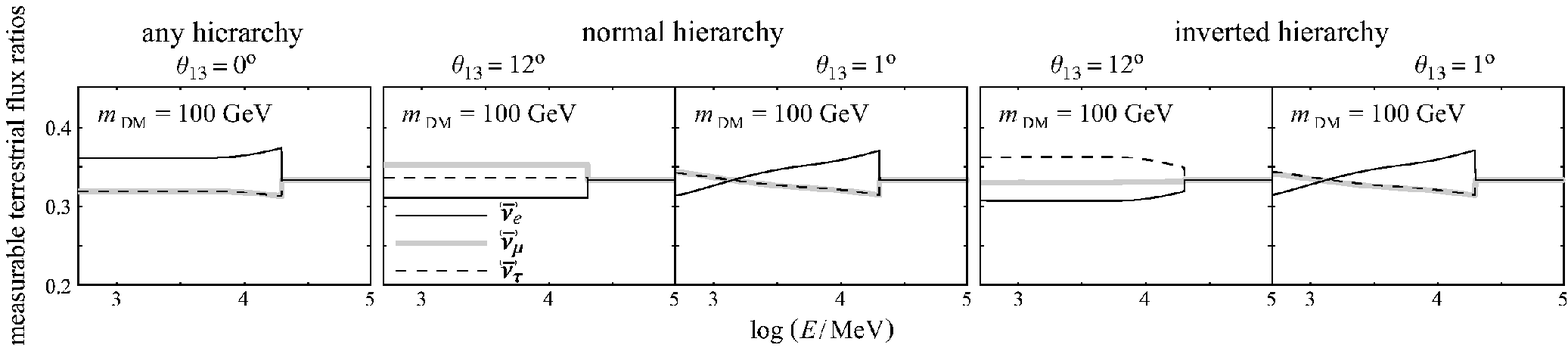}
\end{center}
\vskip-10pt
\caption{
Terrestrial fluxes versus neutrino energy $E$ 
for WIMP annihilation predominantly into $W^+W^-$. 
The plotted observable $(2\Phi_\nu+\Phi_{\overline{\nu}})/3$
and the neutrino parameters 
are the same as those in Figs.\ \ref{fig6}--\ref{fig8}. 
The $\numu$ and $\nutau$ spectra as well as the $\numubar$ and $\nutaubar$ spectra
are interchanged if $\delta=\pi$ is chosen. 
Sensitivities to $\theta_{13}$ and the mass hierarchy 
are comparable to those in Figs.~\ref{fig6}--\ref{fig8}. 
}
\label{fig9} 
\end{figure*}

The solar flavor spectra of Fig.~\ref{fig5} get processed by the physics we have derived 
in this paper.  The resulting flavor spectra at Earth are 
shown in Figs.~\ref{fig6}--\ref{fig9}
for the $b\,\bbar$, $\tau^+\,\tau^-$, 95\%~$b\,\bbar\,+\,$5\%~$\tau^+\,\tau^-$, 
and $W^+\,W^-$ WIMP-annihilation channels, respectively.
For each of the three sample WIMP masses $m_{\rm DM}=10$, 30, and 100~GeV, 
the three values $\theta_{13}=0^\circ$, $1^\circ$, and $12^\circ$ are considered.
We have further taken $\delta=0$, which serves to maximize 
$|\sin\theta_{13}\,\cos\delta |$ and therefore the sensitivity to $\theta_{13}$. 
Results are displayed for the normal and the inverted mass hierarchies.

The greater likelihood of the (I) $W^+\,W^-$
and (II) 95\%~$b\,\bbar\,+\,$5\%~$\tau^+\,\tau^-$ 
decay channels directs our attention to Figs.~\ref{fig8} and \ref{fig9}.
In Fig.~\ref{fig9} one sees the $1:1:1$ flavor ratio at Earth for energies 
above 20~GeV, that results from the democratic production ratio 
in the Sun at these energies for the $W^+\,W^-$ channel.
Below 20~GeV in the $W^+\,W^-$ channel, 
and for the whole energy range in the 95\%~$b\,\bbar\,+\,$5\%~$\tau^+\,\tau^-$ channel,
a comparison of Figs.~\ref{fig8} and \ref{fig9} to Fig.~\ref{fig5} 
makes the point that significant flavor processing 
occurs as neutrinos propagate outward through the Sun.
The flavor processing tends to homogenize the $\mu$ and $\tau$ components.
Nevertheless, in the lower energy region of (I) and higher energy region of 
(II), the flavor ratios differ significantly from unity.

Comparison of Figs.~\ref{fig8} and \ref{fig9} makes the further point that 
flavor ratios arriving at Earth do vary with the WIMP decay channel.
In fact, for the favored channels (I) and (II), flavor signatures are usefully in opposition, 
in that the $W^+\,W^-$ channel offers democratic flavors at higher energies and 
not at lower energies,
whereas the 95\%~$b\,\bbar\,+\,$5\%~$\tau^+\,\tau^-$ channel offers nearly democratic 
flavors at lower energies and not at higher energies.
For the $W^+\,W^-$ channel, the change from non-democratic to democratic is fixed at 
$\sim 20$~GeV, a number traceable to Standard Model masses.
For the 95\%~$b\,\bbar\,+\,$5\%~$\tau^+\,\tau^-$ channel,
the energy for change from nearly democratic to non-democratic depends on the WIMP mass.
For either channel, flavor ratios in the non-democratic regions are seen to depend on 
$\theta_{13}$, and, for $\theta_{13}\neq 0$, on the mass hierarchy.
For the 95\%~$b\,\bbar\,+\,$5\%~$\tau^+\,\tau^-$ channel, 
certain flavor ratios may be as large as 2 at the higher energies;
and in the $W^+\,W^-$ channel, as large as 1.4 below $\sim 20$~GeV.
The flavor ratio of 2 is within experimental reach, 
while ratios of 1.4 or smaller
may present a challenge for some neutrino telescopes~\cite{BBHPWratios}
but perhaps not for others~\cite{BCHHWliqar}.


\section{Summary and Conclusions}
\label{sec:sumNconclude}

The discovery of $\agt$~GeV neutrinos from the Sun would constitute strong evidence
for the annihilation of solar WIMPs.  Competing WIMP models can in principle 
be distinguished by the energy spectrum and flavor ratios of the decay neutrinos 
and antineutrinos.
The flavor ratios detected at Earth will differ from the flavor ratios at injection in the Sun's core.
Due to the decoherence resulting from $\la\,L/E\,\ra$-averaging in the Sun, 
the flavor evolution is described 
by classical probabilities, rather than by quantum-mechanical amplitudes.
This simplifies the analysis.
On the other hand, the evolution is complicated by the matter effects within the Sun.
WIMPS annihilate in a CP-symmetric state, and so the neutrino and antineutrino 
spectra at the point of origin are identical in energies and flavors.
However, since the material of the Sun is matter and not antimatter, 
propagation effects through the Sun are different for 
neutrino versus antineutrino.
In the travel outward from the Sun's core,
two level-crossings are encountered (one if $\theta_{13}=0$), 
each described by an MSW resonance.  
For the normal neutrino-mass hierarchy, both level crossings occur in the neutrino sector;
whereas for the inverted hierarchy, the lower-energy crossing occurs in the neutrino sector 
and the higher-energy resonance occurs in the antineutrino sector.
Thus, the emergent energy and flavor spectra depend also on 
the neutrino-mass hierarchy, as indicated schematically in Fig.~\ref{fig0}.
An additional complication is that evolution at the level crossings may be adiabatic or non-adiabatic, 
depending on solar and neutrino parameters.

In this work, we presented analytic formulas for the complete flavor evolution.  
One virtue of an analytic treatment is the emergence of {\sl post hoc} intuition.
With the gift of hindsight, we therefore were able to present the physics 
of flavor evolution which underlies the analytic formulas.
For example, we were able to derive general statements that \\
(i) If flavor democracy, defined by $1:1:1$ flavor ratios, 
occurs in WIMP decay, then the same $1:1:1$ ratio must be observed on Earth;
in other words, flavor-mixing, even with complicated matter effects, cannot derail democracy! \\
(ii) For maximal $\theta_{32}$ but arbitrary $\theta_{21}$ and $\theta_{13}$,
the replacement $\delta\rightarrow \pi-\delta$ 
interchanges the role of $\nu_{\mu}$ and $\nu_{\tau}$,
and $\numubar$ and $\nutaubar$,
and leaves unaffected the terrestrial $\nue$ and $\nuebar$ fluxes.\\
(iii) For $\theta_{32}$ and $\theta_{12}$ at their tribimaximal values, 
the terrestrial flavor ratios for normal-hierarchy neutrinos 
and inverted-hierarchy antineutrinos 
are independent of $\delta$ to zeroth order in 
$(\delta m^2_{21}/\delta m^2_{32})$.
In addition, the terrestrial $\nu_e$ and $\overline{\nu}_e$ fluxes 
are independent of $\delta$ to zeroth order in 
$(\delta m^2_{21}/\delta m^2_{32})$ for any hierarchy.
(This feature holds only in a finite energy region 
determined by adiabaticity and by $V_eE\gg\delta m^2_{jk}$.)\\
   
Concerning the adiabaticity versus non-adiabaticity, 
we found that level-crossings are adiabatic in the neutrino sector for energies 
up to $\sim 10$~GeV or more in the neutrino sector.  
In the antineutrino sector, there are no level crossing for the normal 
hierarchy and so adiabaticity holds trivially at all energies, 
while for the inverted hierarchy, 
adiabaticity holds up to $\sim 50$~GeV or more.
The energy-onsets of non-adiabatic behavior ($E_{NA}$) are indicated in Figs.~\ref{fig1}-\ref{fig3}.

There is a striking dependence on the value of the unknown mixing-angle $\theta_{13}$ 
in the evolution of flavor away from a single injected flavor in the Sun.
In particular, the limit $\theta_{13}\rarr 0$ is ``discontinuous'' 
in that for $\theta_{13}=0$ the third neutrino mass eigenstate
$|\,3\,\ra$ is completely decoupled from the other two mass eigenstates, 
and there is but a single resonance is encountered in the flavor evolution; 
while for $\theta_{13}\neq 0$, two resonances are encountered,
and even a small nonzero value of $\theta_{13}$ may be amplified by the associated resonance.
In addition, for nonzero $\theta_{13}$, there is sensitivity to 
the CP- and T-symmetric term $\sin\theta_{13}\,\cos\delta$,
where $\delta$ is the CP- and T-violating phase parameter in the unitary mixing-matrix.
Flavor evolution from a single injected flavor in the Sun, for $\theta_{13}$ set to zero, 
is shown in Fig.~\ref{fig1}.
For the maximal allowed value of $\theta_{13}$, 
the same is shown in Fig.~\ref{fig2} for the normal mass-hierarchy, 
and in Fig.~\ref{fig3} for the inverted hierarchy.

We further explored the dependence on $\delta$ in Fig.~\ref{fig4}.
The dependence on $\delta$ enters through the CP- and 
T-even factor $\Re\{U_{e3}\}=\sin\theta_{13}\,\cos\delta$.
To maximize sensitivity to $\delta$ in this figure, 
we have again chosen for $\theta_{13}$ the maximum value allowed 
by the CHOOZ data, $12^\circ$.
The $\mu$ and $\tau$ flavored spectra are sensitive to $\delta$, whereas the $e$ spectrum is not.
The $\numu$ and $\nutau$ spectra and the $\numubar$ and $\nutaubar$ spectra swap among themselves 
for $\delta\rarr\pi-\delta$.
These results are in accord with the general features mentioned above.

WIMP decay does not produce neutrinos of just one flavor.
Consequently, our illustrative evolution results in Figs.~\ref{fig1}-\ref{fig4} do not apply 
directly to the flavor processing of WIMP-decay neutrinos.
Rather, some linear combinations of the single-flavor figures are relevant.
In most WIMP models, the annihilation final states fall into one of two categories.
The two are the $W^+ W^-$ final state and the $95\%,b\bbar+5\%\,\tau^+ \tau^-$ final state.
These Standard Model final states decay further in well known ways 
to produce the neutrinos and antineutrinos emanating from the Sun's core.  
In Fig.~\ref{fig5} we showed the injection neutrino and antineutrino spectra per flavor, 
gleaned from~\cite{Cirelli05}.
A representative sample of WIMP masses were chosen: 10~GeV, 30~GeV, and 100~GeV.
The flavor spectra arriving at Earth, after mixing and matter processing, 
we showed in Figs.~\ref{fig6}-\ref{fig9} 
for various possible  WIMP-annihilation final states.
Each of these Figures is arranged into panels labeled 
by the three values $\theta_{13}=0^\circ,\,1^\circ, 12^\circ$,
and the three values of WIMP mass 10~GeV, 30~GeV, 100~GeV. 
The phase $\delta$ has been set to zero (or $\delta=\pi$ with the interchange
of $\numu$ and $\nutau$ spectra and $\numubar$ and $\nutaubar$ spectra);
we reserve for future work the dependence of the observable flavor spectra on $\delta$~\cite{LWfuture}.  
One feature of the flavor evolution is a tendency to average 
the $\mu$ and $\tau$ flavors. Nevertheless, some interesting nontrivial 
flavor ratios appear in these figures.
 
For the two preferred channels of WIMP annihilation,
it is seen that in the lower energy region of 
the $W^+ W^-$ decay mode, and in the higher energy region 
of the 95\%~$b\,\bbar\,+\,$5\%~$\tau^+\,\tau^-$ decay mode,
flavor ratios differ significantly from unity.
Conversely, in the higher and lower energy regions, respectively,
the flavor ratios are (nearly) democratic.
Since the decay mode implicates the nature of the WIMP,
the experimental ability to distinguish decay modes allows an inference 
of the nature of the WIMP\@.
The energy dependences of the neutrino flavor ratios at Earth are, then, 
the signatures which potentially distinguish the two main competing WIMP models.

The flavor ratios at Earth depend on $\theta_{13}$ and on the neutrino mass hierarchy.
For the purpose of studying the nature of solar WIMPS, it would be very useful if near future 
experiments would better determine these two parameters.
On the other hand, if the nature of the WIMP becomes known from future accelerators such 
as the LHC and/or the ILC, then it becomes conceivable to use the flavor ratios to 
deduce or bound better $\theta_{13}$, and possibly deduce the mass hierarchy.

If one attempts either to infer the nature of the solar WIMP,
or to determine the values of the parameters $\theta_{13}$ and ${\rm sign}(\delta m^2_{32})$,
an algorithm for neutrino flavor processing from the Sun's core outward is essential.
In this paper, we have provided this algorithm.


\acknowledgments
We gratefully acknowledge informative conversations with Dan Hooper and Xerxes Tata.
This work was supported in part by the DOE
under cooperative research agreement No.\ DE-FG02-05ER41360 
and under Grant No.\ DE-FG05-85ER40226.
RL acknowledges support by the European Commission 
under Grant No.\ MOIF-CT-2005-008687,
and TW thanks Manfred Lindner for hospitality and support at MPI-Heidelberg 
during the final stage of paper preparation.

\appendix

\section{Analytic ${\bm U_m}$ for nonzero ${\bm \theta_{13}}$ and ${\bm \delta}$}
\label{matter_mixing} 
In order to obtain a manageable expression for $U_m$, 
we restrict our attention primarily to perturbations about the tribimaximal case \rf{tribimax}. 
In particular, we take $\theta_{12}=\pi/6+\delta\theta_{12}$, 
and we implicitly assume $\tan\theta_{13}\lsim{\cal O}(1)$, 
which is observationally justified. 
On phenomenological grounds, 
we also have $\delta m^2_{21}/\delta m^2_{32}\ll 1$ and $\delta m^2_{jk}/V_e(0) E\ll 1$, 
so that these quantities can serve as additional expansion parameters. 
In the latter relation, 
we have employed the fact 
that our present focus is on GeV neutrinos. 

We begin  
by recalling the defining relation for $U_m$: 
\beq{UmDef} 
H_M=U_m^\dagger\left[U\,\hat{M}\,U^\dagger+\hat{V}\right]U_m\;, 
\eeq 
such that $H_M$ is diagonal. 
For notational brevity we have defined 
$\hat{M}={\rm diag}(-\delta m_{21}^2,0,\delta m_{32}^2)/2E$ (as before),
and $\hat{V}=V_e\,{\rm diag}(1,0,0)$.  
In order to simplify the algebra, 
we decompose $U_m$ as follows: 
\beq{Umdecomp} 
U_m=R_{23}(\theta_{23})\,U^{\dagger}_\delta\,\tilde{U}_m\;,
\eeq 
with $\tilde{U}_m$ to be determined. 
Employing this decomposition and Eq.\ \rf{vacPDG} 
in the definition \rf{UmDef} 
we obtain 
\beq{Umtildedef} 
H_M(\delta)=\tilde{U}^\dagger_m\left[R_{13}U_\delta R_{12}\,\hat{M}\,R_{12}^\dagger 
U_\delta^\dagger R_{13}^\dagger+\hat{V}\right]\tilde{U}_m\;. 
\eeq 
Here, 
we have suppressed the arguments of the rotation matrices, 
and we have used 
that both $R_{23}$ and $U_\delta$ commute with $\hat{V}$. 
Since $\tilde{U}_m$ is unitary 
and diagonalizes 
$R_{13}U_\delta R_{12}\,\hat{M}\,R_{12}^\dagger U_\delta^\dagger 
R_{13}^\dagger+\hat{V}\equiv\tilde{H}$, 
it is composed of normalized eigenvectors of $\tilde{H}$. 
To find these, 
we can employ perturbation theory: 
we have 
$\delta m_{jk}^2\ll EV_e$, 
so that $R_{13}U_\delta R_{12}\,\hat{M}\,R_{12}^\dagger U_\delta^\dagger 
R_{13}^\dagger\equiv\delta\tilde{H}$ 
can be treated as a perturbation when compared to $\hat{V}$. 

The first step in time-independent perturbation theory is 
to identify the zeroth-order eigenstates, 
which in the present situation are those of $\hat{V}$. 
Up to a phase, 
the first such eigenvector clearly is
\beq{ev1}
v_1=(1,0,0)\;. 
\eeq
The remaining two eigenvectors $v_2$ and $v_3$, 
which are a linear combination of $(0,1,0)$ and $(0,0,1)$, 
are associated with degenerate eigenvalues. 
They must therefore be determined 
by diagonalizing $\delta\tilde{H}$ 
restricted to the 2--3 subspace. 
This essentially amounts to finding the eigenvectors of the lower right $2\times2$ block of
\begin{widetext}
\bea{explicit} 
\renewcommand{\arraystretch}{1.5}
\delta\tilde{H} & = & \frac{1}{12E}
\left(
\begin{array}{ccc}
 6 \,\delta m^2_{32} \sin^2\theta_{13}-4 \,\delta m^2_{21} \cos^2\theta_{13} & 2 \sqrt{2}\, e^{i \delta/2} \,\delta m^2_{21} \cos\theta _{13} & (2 \,\delta m^2_{21}+3 \,\delta m^2_{32}) \sin 2 \theta _{13} \\
 2 \sqrt{2}\, e^{-i \delta/2} \,\delta m^2_{21} \cos\theta _{13} & -2 \,\delta m^2_{21} & -2 \sqrt{2}\, e^{-i \delta/2} \,\delta m^2_{21} \sin\theta _{13} \\
 (2 \,\delta m^2_{21}+3 \,\delta m^2_{32}) \sin 2 \theta _{13} & -2 \sqrt{2}\, e^{i \delta/2} \,\delta m^2_{21} \sin\theta _{13} & 6 \,\delta m^2_{32} \cos^2\theta _{13}-4 \,\delta m^2_{21} \sin^2\theta _{13}
\end{array}
\right)\nonumber\\
&&{}+
\frac{\delta m^2_{21}}{6E}
\left(
\begin{array}{lll}
 2 \sqrt{2} \cos ^2\theta _{13} & e^{i \delta/2} \cos\theta _{13} & -\sqrt{2} \sin 2\theta _{13} \\
 e^{-i \delta/2} \cos\theta _{13} & -2 \sqrt{2} & -e^{-i \delta/2} \sin\theta_{13} \\
 -\sqrt{2} \sin 2 \theta _{13} & -e^{i \delta/2} \sin\theta _{13} & 2 \sqrt{2} \sin^2\theta _{13}
\end{array}
\right)\delta\theta_{12}\,.
\eea
\end{widetext}
Here, $\delta\theta_{12}$ denotes deviations of $\theta_{12}$ 
from its tribimaximal value 
implicitly determined by $\sin^2\theta_{12}=1/3$. 
The above expression \rf{explicit}
is valid at first order in $\delta\theta_{12}$. 
We remark 
that the correction term involving $\delta\theta_{12}$ 
contains $\delta m^2_{21}$ as an overall factor, 
but $\delta m^2_{32}$ is absent in the $\delta\theta_{12}$ term. 
This essentially arises because in the definition of $\delta\tilde{H}$ 
the $R_{12}$ matrix mixes only the upper $2\times2$ block of $\hat{M}$, 
so that any occurrence of $\theta_{12}$ 
is necessarily accompanied by a $\delta m^2_{21}$ factor. 

For the remaining zeroth-order eigenvectors, 
we now find at leading order in $\delta m^2_{21}/\delta m^2_{32}$
\beq{ev2} 
\renewcommand{\arraystretch}{1.5}
v_2=
\left(
\begin{array}{c}
0\\
e^{-i\delta/2}\\
\displaystyle\frac{\sqrt{2}}{3}\,\displaystyle\frac{\tan\theta_{13}}{\cos\theta_{13}}\,\displaystyle\frac{\delta m^2_{21}}{\delta m^2_{32}}
\end{array}
\right),
\eeq 
and
\beq{ev3} 
\renewcommand{\arraystretch}{1.5}
v_3=
\left(
\begin{array}{c}
0\\
-\displaystyle\frac{\sqrt{2}}{3}\,\displaystyle\frac{\tan\theta_{13}}{\cos\theta_{13}}\,\displaystyle\frac{\delta m^2_{21}}{\delta m^2_{32}}\,e^{-i\delta/2}\\
1
\end{array}
\right).
\eeq 
We remind the reader 
that normalized eigenvectors are only defined up to phases. 
Note that these expressions are independent of $\delta\theta_{12}$. 
This is a consequence of the fact 
that $\delta m^2_{32}$ is absent in the $\delta\theta_{12}$ term in Eq.\ \rf{explicit}, 
so that the $\delta\theta_{12}$ correction 
contains a further suppression 
relative to the $\delta\theta_{12}$-independent contribution in $\delta\tilde{H}$.

The corrections to these eigenvectors 
are given by the usual perturbation-theory formula 
\beq{corrformula} 
\delta v_j=\sum_{k}\frac{v^\dagger_k\,\delta\tilde{H}\,v_j}{\lambda_j-\lambda_k}\,v_k\;,
\eeq 
where $\lambda_j$ denotes the eigenvalue of $\hat{V}$ 
corresponding to $v_j$, 
i.e., $\lambda_1=V_e$ and $\lambda_2=\lambda_3=0$. 
The sum in Eq.\ \rf{corrformula} runs over all eigenvectors 
that do not belong to the eigenspace containing $v_j$. 
We obtain 
\bea{deltaev} 
\delta v_1&=&
\frac{1}{V_e E}\left(
\renewcommand{\arraystretch}{1.5}
\begin{array}{c}
0\\
\frac{1}{3\sqrt{2}}\, e^{-i \delta/2} \,\delta m^2_{21} \cos\theta _{13}\\
\frac{1}{12}(3 \,\delta m^2_{32}+2\,\delta m^2_{21}) \sin 2 \theta _{13}
\end{array}
\right),\nonumber\\
\delta v_2&=&
\frac{1}{V_e E}\left(
\renewcommand{\arraystretch}{1.5}
\begin{array}{c}
-\frac{1}{3\sqrt{2}}\,\delta m^2_{21}\cos\theta_{13}\\
0\\
0
\end{array}
\right),\nonumber\\
\delta v_3&=&
\frac{1}{V_e E}\left(
\renewcommand{\arraystretch}{1.5}
\begin{array}{c}
-\frac{1}{12}(3 \,\delta m^2_{32}+2\,\delta m^2_{21}) \sin 2 \theta _{13}\\
0\\
0
\end{array}
\right), \phantom{m m}
\eea 
at leading order in $\delta m^2_{21}/\delta m^2_{32}$. 
It follows 
that our approximation $\tilde{U}_m\simeq(v1+\delta v_1,v2+\delta v_2,v3+\delta v_3)$ is explicitly given by
\begin{widetext}
\renewcommand{\arraystretch}{1.5}
\beq{Utilde} 
\tilde{U}_m=
\left(
\begin{array}{ccc}
 1 & -\displaystyle\frac{\delta m^2_{21}\cos \theta _{13} }{3 \sqrt{2} V_e E } & -\displaystyle\frac{(2 \,\delta m^2_{21}+3 \,\delta m^2_{32}) \sin 2\theta _{13}}{12 V_e E } \\
 \displaystyle\frac{\delta m^2_{21}e^{-i\delta/2}\cos \theta _{13} }{3 \sqrt{2} V_e E } & e^{-i\delta/2} &
   -\displaystyle\frac{\sqrt{2}\, e^{-i\delta/2} \, \delta m^2_{21}  \tan \theta _{13}}{3 \,\delta m^2_{32}\cos \theta _{13}} \\
 \displaystyle\frac{(2 \,\delta m^2_{21}+3 \,\delta m^2_{32}) \sin 2\theta _{13}}{12 V_e E } & \displaystyle\frac{\sqrt{2} \,\delta m^2_{21}  \tan \theta _{13}}{3 \,\delta m^2_{32}\cos \theta _{13}} & 1
\end{array}
\right),
\eeq 
\end{widetext}
which is valid to linear order in both $\delta m^2_{32}/V_eE$ and $\delta m^2_{21}/\delta m^2_{32}$. 
First-order terms in $\delta\theta_{12}$ are absent. 
This again arises 
because the $\delta\theta_{12}$ contributions in Eq.\ \rf{explicit}
suffer from an additional suppression 
relative to the leading-order terms in $\delta\tilde{H}$. 
As pointed out in Sec.~\ref{subsec:matteroscillations}, 
the results for antineutrinos 
can be obtained by 
reversing the signs of $V_e$ and $\delta$ 
in the above equations. 

Note that in $\tilde{U}_m$ and thus in $\U2_{\hspace{-.3mm}m}$, 
corrections containing $\theta_{13}$ or $\delta$ 
are also suppressed by $\delta m^2_{21}/\delta m^2_{32}\ll 1$ 
or $\delta m^2_{jk}/V_e(0) E\ll 1$.
This implies 
that for GeV neutrinos at the solar core, 
the matter mixing is largely unaffected 
by $\theta_{13}$ and $\delta$ perturbations 
about the tribimaximal case. 
Significant $\theta_{13}$ and $\delta$ effects 
on the terrestrial fluxes 
can therefore only arise 
through the non-adiabatic transitions described by the matrix $P$
and through the vacuum mixing matrix $U$, 
which determines the flavor content 
of the mass states at the detector. 
 
We remind the reader 
that each column vector in $\tilde{U}_m$ is only defined up to a phase, 
an ambiguity without physical significance. 
The column-vector order in $\tilde{U}_m$, 
on the other hand, 
is to be selected such 
that the diagonal elements of $\tilde{U}_m^\dagger\tilde{H}\tilde{U}_m$ 
exhibit the same ordering as those of the matrix $\hat{M}$. 
The arbitrary ordering choice in Eq.\ \rf{Utilde} gives 
\begin{widetext} 
\beq{Htildediag} 
\tilde{U}_m^\dagger\tilde{H}\tilde{U}_m=
\frac{1}{2E}
\left( 
\begin{array}{ccc} 
 2 V_e E+ \delta m_{32}^2\sin ^2\theta _{13}  & 0 
   & 0 \\
 0 & 0 & 0 \\
 0 & 0 & \delta m_{32}^2 \cos ^2\theta _{13}  
\end{array}
\right)+\cdots\;,
\eeq 
\end{widetext} 
where the ellipsis indicates terms of second or higher order in $\delta m^2_{32}/V_eE$ 
and $\delta m^2_{21}/\delta m^2_{32}$. 
The present range of parameters 
yields $0\ll|3 \cos ^2\theta _{13} \delta m_{32}^2|\ll|6 V_e E|$, 
where the first inequality 
rests on our assumption that $\theta_{13}$ is not too large. 
The correct ordering therefore 
depends on the signs of $V_e$ and $\delta m^2_{32}$. 
It follows 
that the required adjustments can only be implemented 
once particle type ($\nu$ vs.\ $\overline{\nu}$) 
and mass hierarchy are specified. 

Apart from the column-vector order in $\tilde{U}_m$, 
Eq.\ \rf{Umdecomp} can be used to find 
explicit first-order expressions for $U_m$, 
and thus $\U2_{\hspace{-.3mm}m}$. 
The matrix $\U2_{\hspace{-.3mm}m}$ takes the relatively simple form 
\begin{widetext}
\beq{U2m_expression} 
\U2_{\hspace{-.3mm}m}=
\left( 
\begin{array}{ccc} 
 1 & 0 & 0 \\
 0 & \cos^2\theta_{23} & \sin^2\theta_{23} \\
 0 & \sin^2\theta_{23} & \cos^2\theta_{23} 
\end{array}
\right)+
\frac{\sqrt{2}\cos\delta\sin\theta_{13}\sin2\theta_{23}}{3\cos^2\theta_{13}}
\left( 
\begin{array}{rrr} 
 0 & 0 & 0 \\
 0 & 1 & -1 \\
 0 & -1 & 1 
\end{array}
\right)\frac{\delta m^2_{21}}{\delta m^2_{32}}\;.
\eeq 
\end{widetext} 
This expression does not contain ${\cal O}(\delta m^2_{jk}/V_e E)$ and ${\cal O}(\delta \theta_{12})$ terms. 
Only the sign of $V_e$ will eventually play a role, 
as it matters for the column-vector ordering in $\U2_{\hspace{-.3mm}m}$. 
It is again apparent 
that the effects of $\theta_{13}$ and $\delta$  
on the matrix $\U2_{\hspace{-.3mm}m}$ 
are suppressed by more than an order of magnitude 
from the $\delta m^2_{21}/\delta m^2_{32}$ ratio.
For the observable T-conserving matrix elements of $\U2_{\hspace{-.3mm}m}$,
the $\delta$ dependence can only enter in the combination 
$\Re(e^{i\delta}\sin\theta_{13})=\cos\delta\,\sin\theta_{13}$.

\section{Equality of low- and high-energy flavor ratios for ${\bm \theta_{13}=0}$}
\label{accident}

In the low-energy limit $E\to 0$,
matter effects are negligible,
so that $P\,\U2_{\hspace{-.3mm}m}^T\rightarrow\U2^T$ in Eq.\ \rf{matrixform2}. 
In other words, 
the vacuum case \rf{matrixform} arises. 
This general result must hold for all values of the $\theta_{13}$ angle. 
In the high-energy limit $E\to\infty$, 
both $P$ and $\U2_{\hspace{-.3mm}m}$ in Eq.\ \rf{matrixform2} are nontrivial 
because all matter effects must be taken into account. 
It follows 
that the low- and high-energy limits 
are distinguished by the difference in the $\U2^T$ and $P\,\U2_{\hspace{-.3mm}m}^T$ matrices, 
where the latter is to be evaluated for neutrinos (as opposed to antineutrinos)
in the limit $E\rightarrow\infty$. 
We will show that $\U2^T$ and $P\,\U2_{\hspace{-.3mm}m}^T$ are identical 
when $\theta_{13}=0$.

For $\theta_{13}=0$, 
the Hamiltonian can be transformed 
into the form $H'_F$ given by Eq.\ \rf{part_diag1}.
As noted in the discussion leading to Eq.~\rf{dmxprodn},
the off-diagonal pieces of $H'_F$ vanish
when $E$ is taken to approach infinity.
It follows
that $U_m=R_{23}(\theta_{23})$ is one possible matrix
that diagonalizes 
the original Hamiltonian $H_F$
at high energies.
The remaining ambiguity (up to phases) in the construction of $U_m$ 
is determined with our column-vector ordering convention.
To this end, 
we need to fix the mass hierarchy 
and take $\delta m^2_{32}>0$ as an example. 
The reader is invited to verify 
that the inverted hierarchy produces the same final result, 
as expected due to the decoupling of the $|3,r\rangle$ state for $\theta_{13}=0$. 
Note that $V_e(0)$ is positive 
because are considering neutrinos, and not antineutrinos.
It follows 
that the largest eigenvalue in $H'_F$ 
is given by $V_e(0)$ 
located in the upper left entry. 
To match our convention, 
this eigenvalue should appear as bottom right matrix element; 
the remaining two eigenvalues are already in the correct order.
We thus modify our $U_m$ above with an appropriate permutation matrix,
to get
\beq{Udef}
U_m=R_{23}(\theta_{23})\left(\begin{array}{ccc}
0 & 0 & 1\\1 & 0 & 0\\ 0 & 1 & 0
\end{array}\right)
\eeq
as the transformation 
that diagonalizes the high-energy flavor-basis Hamiltonian
for $\theta_{13}=0$.

Next, we turn to the crossing-probability matrix $P$.
For $E\gg\TEna$, 
$P_c^l\rarr\cos^2\theta_{12}$ and $P_c^h\rarr\cos^2\theta_{13}=1$
at the respective lower and higher resonances.
Note that the latter of these probabilities
is consistent with the fact
that the $|3,r\rangle$ state decouples for $\theta_{13}=0$.
Employing these results in Eq.\ \rf{normtransprob}
leads to
\beq{limtransprob}
P=
\left(\begin{array}{ccc}
\sin^2\theta_{12} & \cos^2\theta_{12} & 0\\
\cos^2\theta_{12} & \sin^2\theta_{12} & 0\\
0 & 0 & 1
\end{array}\right)\!\!
\left(\begin{array}{ccc}
1 & 0 & 0\\
0 & 0 & 1\\
0 & 1 & 0
\end{array}\right)
\eeq
for the crossing probabilities
at high energies in the case of a vanishing $\theta_{13}$ angle.

We are now in the position 
to give the simple expression  
\beq{U2Pdef}
{}\hspace{-1.5mm}P\,\U2^T_{\hspace{-.3mm}m}
=\left(\!\!\begin{array}{ccc}
\cos^2\!\theta_{12}\! & \sin^2\!\theta_{12}\cos^2\!\theta_{23}\! & \sin^2\!\theta_{12}\sin^2\!\theta_{23}\\
\sin^2\!\theta_{12}\! & \cos^2\!\theta_{12}\cos^2\!\theta_{23}\! & \cos^2\!\theta_{12}\sin^2\!\theta_{23}\\
0 & \sin^2\!\theta_{23} & \cos^2\!\theta_{23}
\end{array}\!\!\right)
\eeq
for the high-energy limit of the matrix $P\,\U2^T_{\hspace{-.3mm}m}$ for neutrinos.
Comparison of this result with $\U2^T$ then establishes 
that the these two matrices are identical. 
Therefore, the low- and high-energy terrestrial neutrino flavor ratios are also identical. 
Moreover,
the low-energy antineutrino fluxes, which are  
obtained by replacing $U\to U^{*}$ in Eq.\ \rf{probvacuum},
are equal to the above low-energy neutrino results. 
Both of these features 
are apparent in Fig.~\ref{fig1}. 

\section{Proof of Feature (\lowercase{i}) of Sec.~\ref{subsec:delta}}
\label{proof1}

We first note 
that the level-crossing probabilities 
presented in Sec.~\ref{subsec:nonadiabatic}  
act on mass eigenstates, and so are independent of $\delta$. 
It follows that $\delta$ can only affect the terrestrial neutrino fluxes 
via the $\U2$ and $\U2_{\hspace{-.3mm}m}$ flavor-content matrices. 
This implies 
that possible phase effects 
may enter only at the production and detection sites. 
That neutrino propagation 
including non-adiabatic transitions in the Sun 
are unaffected by $\delta$ does not come as a surprise, 
as our jump-probability approximation involves only two levels at each resonance, 
and in two-flavor systems the phase in the mixing matrix can be removed. 

Thus, to show that the replacement $\delta\rightarrow\pi-\delta$ 
leaves unaffected the $\nue$ and $\nuebar$ spectra, while interchanging 
the $\numu$ and $\nutau$ spectra, and $\numubar$ and $\nutaubar$ spectra,
it is sufficient to show that the replacement 
leaves unaffected the first row in both $\U2$ and $\U2_{\hspace{-.3mm}m}$,
but interchanges the lower two rows. 
In mathematical terms, we must show that 
\bea{claim1} 
(\U2)_{e j}(\delta)&=&(\U2)_{e j}(\pi-\delta)\;,\nonumber\\
(\U2)_{\mu j}(\delta)&=&(\U2)_{\tau j}(\pi-\delta)\;,
\eea 
and   
\bea{claim2} 
(\U2_{\hspace{-.3mm}m})_{e j}(\delta)&=&(\U2_{\hspace{-.3mm}m})_{e j}(\pi-\delta)\;,\nonumber\\
(\U2_{\hspace{-.3mm}m})_{\mu j}(\delta)&=&(\U2_{\hspace{-.3mm}m})_{\tau j}(\pi-\delta)\;,
\eea 
for $j=1,2,3$. 

Explicit evaluation of $U(\delta)$ and $U(\pi-\delta)$ 
with $\theta_{32}=45^\circ$ and $\theta_{21}$ and $\theta_{13}$ arbitrary yields
\beq{Urelations}
\renewcommand{\arraystretch}{1.2}
\begin{array}[b]{r@{\;\;=\;\;}lcl}
U_{ej}(\delta) & U_{ej}(\pi-\delta) & \quad & j=1,2\;,\\
U_{ej}(\delta) & -U_{ej}^*(\pi-\delta) & \quad & j=3\;,\\
U_{\mu j}(\delta) & -U_{\tau j}^*(\pi-\delta) & \quad & j=1,2\;,\\
U_{\mu j}(\delta) & U_{\tau j}(\pi-\delta)& \quad & j=3\;. 
\end{array}
\eeq 
These relations imply our condition \rf{claim1}, 
as can be verified 
by multiplying each equation 
with its complex conjugate. 
We remark that this result already establishes Feature~(i) 
for cases with negligible matter effects, such as in vacuum. 

Large matter potentials $V_e(r)$
can definitely modify neutrino flavor fluxes, 
as we have seen in the text. 
However, 
the $\nu_\mu$--$\nu_\tau$ relationship is left unaffected by $V_e(r)$ 
because only $\nu_e$ interacts with the solar background electrons. 
The $\nu_\mu$--$\nu_\tau$ part of the above vacuum result, 
namely that the $\nu_\mu$ and $\nu_\tau$ fluxes switch 
under $\delta\rightarrow\pi-\delta$, 
should therefore generalize to situations with matter. 

To see this explicitly 
and to establish furthermore 
that the $\nu_e$ flux remains unchanged, 
we need to prove condition \rf{claim2}. 
The mixing matrix in matter 
is defined via Eq.\ \rf{UmDef}. 
With the $\delta$ dependence displayed, 
this definition reads for $U_m(\delta)$: 
\beq{UmDef2} 
H_M(\delta)=U_m(\delta)^\dagger\left[U(\delta)\,\hat{M}\,U^\dagger(\delta)+\hat{V}\right]U_m(\delta)\;. 
\eeq 
In order to relate $U_m(\pi-\delta)$ and $U_m(\delta)$, 
we start from the analogous equation for $U_m(\pi-\delta)$, 
i.e., 
$H_M(\pi-\delta)=U_m^\dagger(\pi-\delta)\left[U(\pi-\delta)\,
   \hat{M}\,U^\dagger(\pi-\delta)+\hat{V}\right]U_m(\pi-\delta)$, 
and implement the symmetries \rf{Urelations}. 
In matrix form, 
these symmetries may be rewritten as 
\beq{Usymmetries} 
U(\pi-\delta)=F\,U^*(\delta)\,S\;, 
\eeq 
where the matrix $F$ interchanges the $\mu$ and $\tau$ flavors, 
and the matrices $S$ and $F$ change various signs: 
\bea{SFdef} 
F & = & \left(\begin{array}{rrr}
 -1 &  0 & 0   \\
 0 &  0 & 1 \\
 0 &  1 & 0
\end{array}\right)\;,\nonumber\\
S & = & \left(\begin{array}{rrr}
 -1 &  0 & 0   \\
 0 &  -1 & 0 \\
 0 &  0 & 1
\end{array}\right)\;.
\eea 
The specific expressions for the matrices $\hat{M}$, $\hat{V}$, $F$, and $S$ 
are such that $S\hat{M}S^\dagger=\hat{M}$ and $F \hat{V}F^\dagger=\hat{V}$. 
With these observations, 
the defining relation for $U_m(\pi-\delta)$ can be cast into the following form: 
$H_M(\pi-\delta)=U_m^\dagger(\pi-\delta)\,F\left[U^*(\delta)\,\hat{M}\,U^T(\delta)+\hat{V}\right]F^\dagger\, U_m(\pi-\delta)$. 
Recalling 
that $H_M(\pi-\delta)$, $\hat{M}$, $\hat{V}$, and $F=F^\dagger$ are real, 
complex conjugation of this equation yields: 
\bea{DefResult} 
\lefteqn{{}\hspace{-8mm}H_M(\pi-\delta)=}\nonumber\\
&& {}\hspace{-10mm}U_m^T(\pi-\delta)\,F\left[U(\delta)\,\hat{M}\,U^\dagger(\delta)+\hat{V}\right]F^\dagger\, U_m^*(\pi-\delta)\;.
\eea 
By definition, 
both $H_M(\pi-\delta)$ and $H_M(\delta)$ are diagonal. 
Moreover, 
they exhibit the same ordering of eigenvalues 
because they are associated with the same mass hierarchy and particle type ($\nu$ vs.\ $\overline{\nu}$). 
Note also 
that $F^\dagger\,U_m^*(\pi-\delta)$ is unitary. 
These facts 
and comparison of Eq.~\rf{DefResult} with Eq.~\rf{UmDef2} show 
that $F^\dagger\,U_m^*(\pi-\delta)$ satisfies the defining relation~\rf{UmDef2}
for $U_m(\delta)$. 
We thus find 
\beq{UmRelation} 
U_m(\delta)= 
\left(\begin{array}{ccc}
 e^{i\alpha} &  0 & 0   \\
 0 &  e^{i\beta} & 0 \\
 0 &  0 & e^{i\gamma}
\end{array}\right)
F^\dagger\,U_m^*(\pi-\delta)\;. 
\eeq 
Here, 
the three undetermined phases $\alpha$, $\beta$, and $\gamma$ 
are unobservable, 
but are included for completeness. 
The condition \rf{claim2} 
follows now directly from~\rf{UmRelation}.
Feature (i) is thereby established.

\section{Proof of Feature (\lowercase{ii}) of Sec.~\ref{subsec:delta}}
\label{proof2}

The general terrestrial flavor fluxes 
are given by Eq.\ \rf{matrixform2}. 
The matrix $A\equiv\U2\,P\,\U2_{\hspace{-.3mm}m}^T$ in this equation 
can explicitly be obtained at first order 
with the result \rf{U2m_expression}. 
At leading order in the three small parameters 
$\delta m^2_{21}/\delta m^2_{32}$, $\delta\theta_{12}$, and $\delta\theta_{23}$, 
the generic form of $A$ is 
\beq{generic} 
A=A_0+A_m\frac{\delta m^2_{21}}{\delta m^2_{32}}
+A_{12}\:\delta\theta_{12}+A_{23}\:\delta\theta_{23} \;.
\eeq 
Here, 
$\delta\theta_{12}$ and $\delta\theta_{23}$ 
denote deviations from the tribimaximal values of $\theta_{12}$ and $\theta_{23}$, 
respectively.
Note that terms of order $\delta m^2_{jk}/V_e E$ are absent 
because they do not enter into $\U2_{\hspace{-.3mm}m}$ 
in Eq.~\rf{U2m_expression}, 
and $U$ in vacuum depends only on the mixing angles. 
Below we give explicit expressions for the matrices 
$A_0$ , $A_m$, $A_{12}$, and $A_{23}$.
These explicit expressions require the knowledge of the column-vector ordering 
in Eq.\ \rf{U2m_expression}.
Thus, the specification of particle type ($\nu$ vs.\ $\overline{\nu}$) 
and the mass hierarchy is necessary.

We seek to prove that in the adiabatic limit ($P=\unitmatrix$),
for NH neutrinos and IH antineutrinos, that $A_0$ is independent of $\delta$;
and that with either mass hierarchy, the first row of $A_0$ is independent of 
$\delta$.  When $A_0$ is independent of $\delta$, then all flavor spectra 
are independent of $\delta$ at leading order; when the first row of $A_0$ 
is independent of $\delta$, then the $\nue$ and $\nuebar$ spectra are 
independent of $\delta$ at leading order.
Below we determine the full $\theta_{13}$ and $\delta$ dependence of $A_0$.
For completeness, we also determine $A_{12}$ and $A_{23}$ 
at zeroth order in the small parameters. 
It turns out that 
$A_{m}=0+{\cal O}(\theta_{13},\delta\theta_{12},\delta\theta_{23})$ in all cases,
and so we do not consider it any further. 

To estimate the effects of non-adiabaticity ($P\neq\unitmatrix$),
we also consider the case 
of a completely non-adiabatic higher-energy resonance 
while maintaining adiabaticity at the lower-energy resonance. 
This requires that $\theta_{13}$ be small enough such  
that the non-adiabaticity at the higher-energy resonance occurs 
at energies where the lower-energy resonance is still adiabatic. 

{\bf Neutrinos within the normal hierarchy.} 
This case is characterized by $0<\delta m^2_{32}<V_e E$. 
Inspection of Eq.\ \rf{Htildediag} shows 
that the $\tilde{U}_m$ column vectors in Eq.\ \rf{Utilde} 
need to be rearranged according to 
$1 \rightarrow 3 \rightarrow 2 \rightarrow 1$, i.e., 
\beq{Utilde_neut_normal} 
\tilde{U}_m \rightarrow \tilde{U}_m 
\left(
\begin{array}{ccc}
 0 & 0 & 1 \\
 1 & 0 & 0 \\
 0 & 1 & 0
\end{array}
\right). 
\eeq 
This gives the following explicit form 
of the $A_0$, $A_{12}$, and $A_{23}$ matrices: 
\beq{R0_neut_normal} 
A_0=\frac{1}{4}
\left(
\begin{array}{ccc}
 4 s^2_{13} & 2 c^2_{13} & 2 c^2_{13} \\
 2 c^2_{13} & 1+s^2_{13} & 1+s^2_{13} \\
 2 c^2_{13} & 1+s^2_{13} & 1+s^2_{13}
\end{array}
\right),
\eeq 
\beq{A12_neut_normal} 
A_{12}=0\;,
\eeq 
\beq{A23_neut_normal} 
A_{23}=
\frac{1}{3}\left(
\begin{array}{rrr}
 0 & -1  & 1 \\
 3 & -1 & -2 \\
 -3 & 2 & 1
\end{array}
\right).
\eeq 
Inspection of Eq.\ \rf{R0_neut_normal} shows 
that $\delta$ leaves unaffected the terrestrial flavor fluxes at leading order, 
which establishes Feature (ii) 
for normal-hierarchy neutrinos. 
Moreover, 
the second and third rows of $A_0$ are identical, 
implying degeneracy between the $\nu_\mu$ and $\nu_\tau$ fluxes 
at this order.

The results \rf{R0_neut_normal}, \rf{A12_neut_normal}, and \rf{A23_neut_normal} 
become invalid for small $\theta_{13}$. 
In this situation, 
the resonance involving $|3,r\rangle$ 
fails to meet the above adiabaticity condition $P=\unitmatrix$: 
the onset of non-adiabaticity \rf{onset} 
applies here with $\theta=\theta_{13}$, 
which lies below the resonance energy \rf{highres} 
for $\theta_{13}$ small enough. 
This non-adiabaticity 
is associated with the decoupling of the state $|3,r\rangle$, 
as discussed in Sec.\ \ref{subsec:diagonalizing}. 
We display the $A$-matrices for a completely adiabatic lower-energy resonance 
and a completely non-adiabatic higher-energy resonance: 
\begin{widetext} 
\beq{R0_neut_normal_na} 
A_0=\frac{1}{6}
\renewcommand{\arraystretch}{1.5}
\left(
\begin{array}{ccc}
 2 c^2_{13} & 2+s^2_{13} & 2+s^2_{13} \\
 2+s^2_{13}-\sqrt{8} \,c_\delta \,s_{13} & 2-\frac{1}{2} s^2_{13}+\sqrt{2} \,c_\delta \,s_{13} &
   2-\frac{1}{2} s^2_{13}+\sqrt{2} \,c_\delta \,s_{13} \\
 2+s^2_{13}+\sqrt{8} \,c_\delta \,s_{13} & 2-\frac{1}{2} s^2_{13}-\sqrt{2} \,c_\delta \,s_{13} &
   2-\frac{1}{2} s^2_{13}-\sqrt{2} \,c_\delta \,s_{13}
\end{array}
\right),
\eeq 
\end{widetext} 
\beq{A12_neut_normal_na} 
A_{12}=\frac{1}{3\sqrt{2}}\left(
\begin{array}{rrr}
 4 & -2  & -2 \\
 -2 & 1 & 1 \\
 -2 & 1 & 1
\end{array}
\right),
\eeq 
\beq{A23_neut_normal_na} 
A_{23}=
\frac{2}{3}\left(
\begin{array}{rrr}
 0 & -1  & 1 \\
 -1 & 1 & 0 \\
 1 & 0 & -1
\end{array}
\right).
\eeq 
In this non-adiabatic case, only the first row of $A_0$ is 
independent of $\delta$, and so only the $\nue$ and $\nuebar$ 
spectra are independent of $\delta$ at lowest order.

{\bf Antineutrinos within the normal hierarchy.}  
As the sign of $V_e$ for antineutrinos 
is opposite that for neutrinos, 
we have now $V_e E<0<\delta m^2_{32}$. 
It follows 
that the diagonal entries of $\tilde{U}_m^\dagger\tilde{H}\tilde{U}_m$ 
in Eq.\ \rf{Htildediag} 
are correctly ordered already, 
and adjustments in $\tilde{U}_m$ are unnecessary. 
But for antineutrinos we need $\U2_{\hspace{-.3mm}m}^*$ and $\U2^*$, 
which can be obtained 
by simply changing the sign of $\delta$ 
in the final expressions for $A_0$, $A_{12}$, and $A_{23}$. 
This yields
\begin{widetext} 
\beq{R0_anti_normal} 
A_0=\frac{1}{6}
\renewcommand{\arraystretch}{1.5}
\left(
\begin{array}{ccc}
 4 c^2_{13} & 1+2 s^2_{13} & 1+2 s^2_{13} \\
 3-2 c^2_{13}+\sqrt{8} \,c_\delta \,s_{13} & \frac{3}{2}+ c^2_{13}- \sqrt{2} \,c_\delta \,s_{13}
  & \frac{3}{2}+ c^2_{13}- \sqrt{2} \,c_\delta \,s_{13} \\
 3-2 c^2_{13}-\sqrt{8} \,c_\delta \,s_{13} & \frac{3}{2}+ c^2_{13}+ \sqrt{2} \,c_\delta \,s_{13} 
 & \frac{3}{2}+ c^2_{13}+\sqrt{2} \,c_\delta \,s_{13}
\end{array}
\right),
\eeq 
\end{widetext} 
\beq{A12_anti_normal} 
A_{12}=\frac{1}{3\sqrt{2}}\left(
\begin{array}{rrr}
 4 & -2  & -2 \\
 -2 & 1 & 1 \\
 -2 & 1 & 1
\end{array}
\right),
\eeq 
\beq{A23_anti_normal} 
A_{23}=
\frac{1}{3}\left(
\begin{array}{rrr}
 0 & -1  & 1 \\
 -1 & 1 & 0 \\
 1 & 0 & -1
\end{array}
\right).
\eeq 
Note 
that there are unsuppressed phase effects, 
since $A_0$ depends on $\delta$. 
However, the first row of $A_0$ remains $\delta$-independent.

We remind the reader 
that resonances are absent in this case. 
Adiabaticity is therefore ensured trivially, 
and the results \rf{R0_anti_normal}, \rf{A12_anti_normal}, and \rf{A23_anti_normal} 
remain valid in the limit $\theta_{13}\to 0$. 
 
{\bf Neutrinos within the inverted hierarchy.} 
This case is characterized 
by $\delta m^2_{32}<0<V_e E$. 
Inspection of Eq.\ \rf{Htildediag} and $\hat{M}$ establishes 
that we have to interchange 
the first and second column in $\tilde{U}_m$, 
which can be implemented via the replacement 
\beq{Utilde_neut_inv} 
\tilde{U}_m \rightarrow \tilde{U}_m 
\left(
\begin{array}{ccc}
 0 & 1 & 0 \\
 1 & 0 & 0 \\
 0 & 0 & 1
\end{array}
\right). 
\eeq 
We then obtain the following expressions for $A_0$, $A_{12}$, and $A_{23}$:
\begin{widetext} 
\beq{R0_neu_inv} 
A_0=
\frac{1}{6}
\renewcommand{\arraystretch}{1.5}
\left(
\begin{array}{ccc}
 2-2 s^2_{13} & 2+s^2_{13} & 2+s^2_{13} \\
 2+s^2_{13}- \sqrt{8} \,c_\delta \,s_{13} & 2-\frac{1}{2} s^2_{13}+\sqrt{2} \,c_\delta \,s_{13} &
   2-\frac{1}{2} s^2_{13}+\sqrt{2} \,c_\delta \,s_{13} \\
 2+s^2_{13}+\sqrt{8} \,c_\delta \,s_{13} & 2-\frac{1}{2} s^2_{13}-\sqrt{2} \,c_\delta \,s_{13} &
   2-\frac{1}{2} s^2_{13}-\sqrt{2} \,c_\delta \,s_{13}
\end{array}
\right),
\eeq 
\end{widetext} 
\beq{A12_neu_inv} 
A_{12}=\frac{1}{3\sqrt{2}}\left(
\begin{array}{rrr}
 4 & -2  & -2 \\
 -2 & 1 & 1 \\
 -2 & 1 & 1
\end{array}
\right),
\eeq 
\beq{A23_neu_inv} 
A_{23}=
\frac{2}{3}\left(
\begin{array}{rrr}
 0 & -1  & 1 \\
 -1 & 1 & 0 \\
 1 & 0 & -1
\end{array}
\right).
\eeq 
It is apparent 
that phase effects are again unsuppressed  
because $A_0$ depends on $\delta$. 
However, the first row of $A_0$ remains $\delta$-independent.

Although a resonance is present in the neutrino sector 
it does not involve the state $|3,r\rangle$. 
Hence, 
the adiabaticity condition is met, 
and the results \rf{R0_neu_inv}, \rf{A12_neu_inv}, and \rf{A23_neu_inv}
keep their validity for vanishing $\theta_{13}$. 

{\bf Antineutrinos within the inverted hierarchy.} 
In this case, we have
$V_e E<\delta m^2_{32}<0$. 
Comparison of the eigenvalue ordering in Eq.\ \rf{Htildediag} 
with that in $\hat{M}={\rm diag}(-\delta m_{21}^2,0,\delta m_{32}^2)/2E$  
shows that we have to interchange 
the first and third column in $\tilde{U}_m$, 
which can be implemented via the replacement 
\beq{Utilde_anti_inv} 
\tilde{U}_m \rightarrow \tilde{U}_m 
\left(
\begin{array}{ccc}
 0 & 0 & 1 \\
 0 & 1 & 0 \\
 1 & 0 & 0
\end{array}
\right). 
\eeq 
The final results for $A_0$, $A_{12}$, and $A_{23}$ 
are similar to those for neutrinos within the normal hierarchy: 
\beq{R0_anti_inv} 
A_0=\frac{1}{4}
\left(
\begin{array}{ccc}
 4 s^2_{13} & 2 c^2_{13} & 2 c^2_{13} \\
 2 c^2_{13} & 1+s^2_{13} & 1+s^2_{13} \\
 2 c^2_{13} & 1+s^2_{13} & 1+s^2_{13}
\end{array}
\right),
\eeq 
\beq{A12_anti_inv} 
A_{12}=0\;,
\eeq 
\beq{A23_anti_inv} 
A_{23}=
\frac{1}{3}\left(
\begin{array}{rrr}
 0 & 1  & -1 \\
 3 & -2 & -1 \\
 -3 & 1 & 2
\end{array}
\right).
\eeq 
We see 
that the present $A_0$ is 
identical to that in Eq.\ \rf{R0_neut_normal}. 
It follows 
that within the adiabatic regime, 
the net behavior 
of IH antineutrinos 
equals that of NH neutrinos 
within the approximations made in this appendix. 
Feature (ii) therefore also holds 
in the present case. 

Let us mention an additional aspect 
shared by NH neutrinos and IH antineutrinos: 
the presence of the resonance involving $|3,r\rangle$. 
Hence, 
the results \rf{R0_anti_inv}, \rf{A12_anti_inv}, and \rf{A23_anti_inv} become invalid 
for small $\theta_{13}$, 
paralleling the above argument for NH neutrinos. 
Finally, we quote the result 
for the completely non-adiabatic limit at the higher-energy resonance: 
\begin{widetext} 
\beq{R0_anti_inv_na} 
A_0=\frac{1}{6}
\renewcommand{\arraystretch}{1.5}
\left(
\begin{array}{ccc}
 4 c^2_{13} & 1+2 s^2_{13} & 1+2 s^2_{13} \\
 3-2 c^2_{13}+ \sqrt{8} \,c_\delta \,s_{13} & \frac{3}{2}+ c^2_{13}- \sqrt{2} \,c_\delta \,s_{13}
  & \frac{3}{2}+ c^2_{13}- \sqrt{2} \,c_\delta \,s_{13} \\
 3-2 c^2_{13}- \sqrt{8} \,c_\delta \,s_{13} & \frac{3}{2}+ c^2_{13}+ \sqrt{2} \,c_\delta \,s_{13} 
 & \frac{3}{2}+ c^2_{13}+\sqrt{2} \,c_\delta \,s_{13}
\end{array}
\right),
\eeq 
\end{widetext} 
\beq{A12_anti_inv_na} 
A_{12}=\frac{1}{3\sqrt{2}}\left(
\begin{array}{rrr}
 4 & -2  & -2 \\
 -2 & 1 & 1 \\
 -2 & 1 & 1
\end{array}
\right)
\eeq 
\beq{A23_anti_inv_na} 
A_{23}=
\frac{1}{3}\left(
\begin{array}{rrr}
 0 & -1  & 1 \\
 -1 & 1 & 0 \\
 1 & 0 & -1
\end{array}
\right).
\eeq

{\bf Summary -- } 
To summarize this appendix, we have shown that when $\theta_{32}$ and $\theta_{21}$ assume their 
tribimaximal values,
then to zeroth order in $\delta m^2_{21}/\delta m^2_{32}$ and $\delta m^2_{32}/V_e(0)\,E$,
one has that \\
(i) the $\nue$ and $\nuebar$ spectra are independent of the phase-parameter $\delta$;\\
(ii)the normal-hierarchy $\numu$ and $\nutau$ spectra are independent of $\delta$; and\\
(iii) the inverted-hierarchy $\numubar$ and $\nutaubar$ spectra are independent of $\delta$.\\
Only the normal-hierarchy $\numubar$ and $\nutaubar$ spectra, and the 
inverted-hierarchy $\numu$ and $\nutau$ spectra depend on $\delta$ in zeroth order.
Finally, we remark that the dependence on $\delta$ in the derived expressions 
has the form $\sin\theta_{13}\cos\delta$, 
as required for CP- and T-invariant observables.


\end{document}